\documentclass[pdflatex,sn-mathphys-num]{sn-jnl}% Math and Physical Sciences Numbered Reference Style 
\usepackage{graphicx}%
\usepackage{multirow}%
\usepackage{amsmath,amssymb,amsfonts}%
\usepackage{amsthm}%
\usepackage{mathrsfs}%
\usepackage[title]{appendix}%
\usepackage{xcolor}%
\usepackage{textcomp}%
\usepackage{manyfoot}%
\usepackage{booktabs}%
\usepackage{algorithm}%
\usepackage{algorithmicx}%
\usepackage{algpseudocode}%
\usepackage{listings}%
\usepackage{mathtools}

\usepackage{array}
\theoremstyle{thmstyleone}%
\theoremstyle{thmstyletwo}%

\theoremstyle{thmstylethree}%

\begin{document}

% Don't do words count from now on!
%TC:ignore

\title[Article Title]{Scale-PINN: Learning Efficient Physics-Informed Neural Networks Through Sequential Correction}
%\title[Article Title]{Scale-PINN: Sequential Correction Algorithm for Learning Efficient Physics-Informed Neural Network Models}


\author[1]{\fnm{Pao-Hsiung} \sur{Chiu}}\email{chiuph@a-star.edu.sg}
\equalcont{These authors contributed equally to this work.}

\author*[1]{\fnm{Jian Cheng} \sur{Wong}}\email{wongj@a-star.edu.sg}
\equalcont{These authors contributed equally to this work.}

\author[1,2]{\fnm{Chin Chun} \sur{Ooi}}\email{ooicc@a-star.edu.sg}

\author[3]{\fnm{Chang} \sur{Wei}}\email{wei\_chang@tju.edu.cn}
\author[3]{\fnm{Yuchen} \sur{Fan}}\email{fanyuchen@tju.edu.cn}

\author[2,4]{\fnm{Yew-Soon} \sur{Ong}}\email{asysong@ntu.edu.sg}

\affil[1]{\orgdiv{Institute of High Performance Computing}, \orgname{A*STAR}, \country{Singapore}}
\affil[2]{\orgdiv{Centre for Frontier AI Research}, \orgname{A*STAR}, \country{Singapore}}
\affil[3]{\orgdiv{School of Mechanical Engineering}, \orgname{Tianjin University}, \country{China}}
\affil[4]{\orgdiv{College of Computing and Data Science}, \orgname{Nanyang Technological University}, \country{Singapore}}


\abstract{Physics-informed neural networks (PINNs) have emerged as a promising mesh-free paradigm for solving partial differential equations, yet adoption in science and engineering is limited by slow training and modest accuracy relative to modern numerical solvers. We introduce the Sequential Correction Algorithm for Learning Efficient PINN (Scale-PINN), a learning strategy that bridges modern physics-informed learning with numerical algorithms. Scale-PINN incorporates the iterative residual-correction principle, a cornerstone of numerical solvers, directly into the loss formulation, marking a paradigm shift in how PINN losses can be conceived and constructed. This integration enables Scale-PINN to achieve unprecedented convergence speed across PDE problems from different physics domain, including reducing training time on a challenging fluid-dynamics problem for state-of-the-art PINN from hours to sub-2 minutes while maintaining superior accuracy, and enabling application to representative problems in aerodynamics and urban science. By uniting the rigor of numerical methods with the flexibility of deep learning, Scale-PINN marks a significant leap toward the practical adoption of PINNs in science and engineering through scalable, physics-informed learning. Codes are available at \url{https://github.com/chiuph/SCALE-PINN}.}

%%================================%%
%% Sample for structured abstract %%
%%================================%%

%Main text – up to 3,500 words, excluding abstract, Methods, references and figure legends.
%Abstract – up to 150 words, unreferenced. 
%Display items – up to 6 items (figures and/or tables). 
%Article should be divided as follows: 
%Introduction (without heading) 
%Results
%Discussion
%Online Methods.
%Results and Methods should be divided by topical subheadings; the Discussion does not contain subheadings.
%References – as a guideline, we typically recommend up to 50.
%Please include tables at the end of your text document. 




\maketitle

%TC:endignore
% Start words count from now on!

\section{Introduction}\label{sec:intro}

%\textit{The Introduction section, of referenced text \cite{bib1} expands on the background of the work (some overlap with the Abstract is acceptable). The introduction should not include subheadings.}

Physics-Informed Neural Networks (PINNs) have emerged as a promising paradigm for solving partial differential equations (PDEs) by embedding governing physics into the training loss. Their mesh-free nature, ability to integrate physics with sparse data, and suitability for inverse problems have sparked widespread interest across computational science, from fluid and solid mechanics to electromagnetism, optics, earth sciences, materials science, electrochemistry, and epidemiology, underscoring their growing prominence as an alternative to numerical simulation methods~\cite{karniadakis2021physics, park2025unifying, tang2022physics, okazaki2022physics, raabe2023accelerating, wang2024physics, kharazmi2021identifiability}.

%Despite their promise, PINNs often face challenges in converging to accurate solutions during training. Issues such as spectral bias and complex loss landscapes can arise during training and impede convergence, particularly for stiff, multi-scale, and high-order PDEs. Efforts to overcome the training challenges of PINNs have led to various methodological advancements, including adaptive sampling, adaptive loss weighting, improved neural architectures, transfer and curriculum learning, and meta-learning. Multiple strategies can be integrated to enhance model performance. For instance, JAX-PI~\cite{wang2023experts} employs Fourier feature embeddings and random weight factorization within its network architecture, alongside training techniques such as causal training, curriculum learning, and adaptive loss weighting. Building on this foundation, PirateNets~\cite{wang2024piratenets} and SOAP~\cite{wang2025gradient} further extend JAX-PI by incorporating a more effective network architecture and initialization scheme with transfer learning and second order optimization. Both models exemplify the potential of PINNs to achieve high solution accuracy across diverse PDE problems.% Yet, despite these advancements, PINNs still lag behind classical numerical solvers in computational efficiency and precision.

However, a key limitation of PINNs thus far, even when employing improved neural architectures and state-of-the-art learning strategies~\cite{liu2025automatic, zhou2025automated, wang2023experts, wang2024piratenets, wang2025gradient}, is their high computational cost-accuracy trade-off. First, the use of dense training samples and large batch sizes improves accuracy but significantly increase computational demands. Second, adaptive sampling and neural tangent kernel-based adaptive loss weighting methods introduce considerable overhead to overcome spectral bias and complex loss landscape issues. Third, the need for curriculum and sequence-to-sequence training to gradually refine the solution from an initial guess to prevent premature convergence is very costly in more complex problems, such as stiff and multi-scale PDEs. Fourth, some studies have turned to second-order optimization methods, including BFGS algorithm and Quasi-Newton variants, to improve accuracy through more precise updates. All of them substantially increase the computational costs as a trade-off for achieving quality outcome. As a result, PINN training is much slower than numerical solvers, limiting their broader adoption in real-world scientific and engineering problems~\cite{mcgreivy2024weak}. Curiously, these strategies derive primarily from the general machine learning literature and do not leverage insights from scientific computing.

%We ask an important question about the future of PINNs for scientific simulations: Would PINNs achieve the \textit{speed (efficiency)} and \textit{precision (reliability)} of modern numerical solvers, in any way?

In seeking to overcome this limitation, we recognize that the scientific computing community has developed a wealth of knowledge and algorithmic techniques over the past decades for efficiently solving complex PDEs---many of which can be potentially adopted to advance PINN methodologies. For example, realizing that lack of neighborhood dependencies information may lead to training failure, numerical differentiation techniques, including the finite difference, finite volume and finite element discretizations, have been employed to successfully enhance physics-informed learning by enforcing local spatial coupling~\cite{bib:Jiang23, bib:Zou24, bib:Roy24, bib:Xiao24, bib:yan2025finite, bib:yamazaki2025finite}. Furthermore, combining automatic differentiation and numerical discretization for PINNs has proven effective in improving both accuracy and computational efficiency~\cite{bib:Chiu22, bib:Wong23}. The artificial eddy viscosity and pseudo-time stepping methods are other notable examples for improving training stability~\cite{bib:Wang23, bib:Cao23, bib:Cao24}. These early successes highlight a promising direction of integrating algorithmic insights from scientific computing into PINN for designing practical and scalable physics-informed learning frameworks. %Yet, many research avenues in this direction remain unexplored, offering opportunities for further innovation and improvement.

While these developments demonstrate the power of infusing discretization wisdom into the PINN framework, they have largely drawn from only one of the two pillars of numerical simulation: discretization and iterative method. Modern numerical methods not only discretize governing equations but also solve the resulting linear systems through carefully designed iterative schemes that guarantee convergence, stability, and precision. Building on this foundation, we demonstrate that the second pillar---iterative schemes---can be a rich source of algorithmic insight for efficient physics-informed learning. In this work, we establish that iterative residual correction, which is a principle at the core of many numerical solvers, can be explicitly realized within the PINN loss formulation with remarkable gains in PINN learning.

\begin{figure*}[h]
\centering
\includegraphics[width=1.\textwidth]{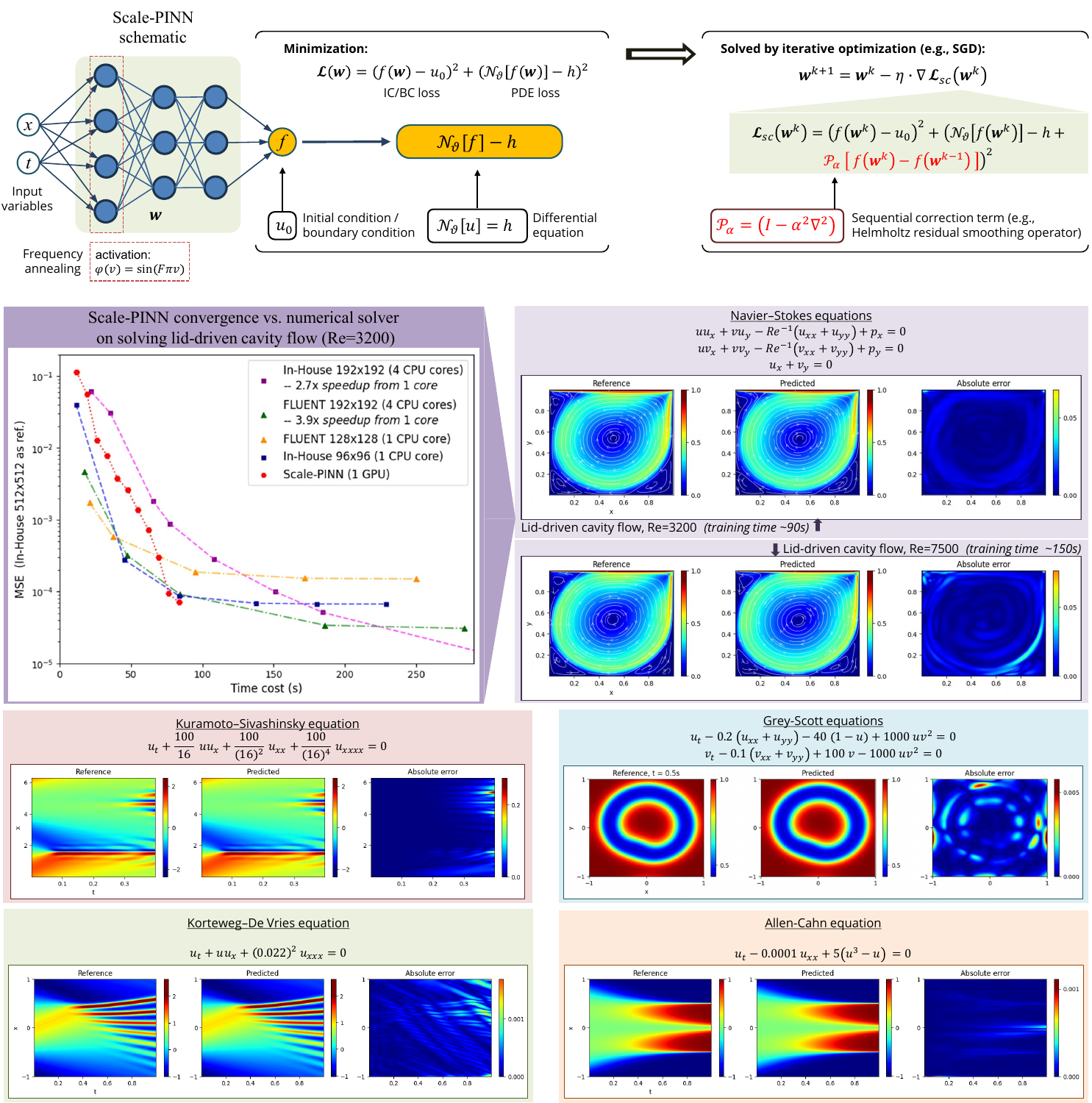}
\vspace*{-3mm}
\caption{Scale-PINN schematic and result highlights. Scale-PINN includes a sequential correction term through application of the residual smoothing operator $\mathcal{P}_{\alpha} = (I - {\alpha^2} \nabla^2 )$ to the change in solution $\mathcal{F} \coloneqq f(\cdot;\boldsymbol{w}^{k})-f(\cdot;\boldsymbol{w}^{k-1})$ during iterative optimization. A convergence plot on the Navier-Stokes (N-S) example, lid‑driven cavity flow at $Re=3200$, shows competitive time‑to‑accuracy versus numerical solvers. Compared to other PINN methods, Scale-PINN solves the lid‑driven cavity flow to state-of-the-art accuracy with unprecedented speed, i.e., $\sim$90s for $Re=3200$ and $\sim$150s for $Re=7500$. Results for Kuramoto–Sivashinsky (K-S), Grey–Scott (G-S), Korteweg–De Vries (KdV), and Allen–Cahn (AC) equations demonstrate accuracy across diverse dynamics. Scale-PINN model architecture and training strategies are detailed in Method~\ref{sec:PINN-full}.}
\label{fig:schemetic}
\end{figure*}

Our proposed Sequential Correction Algorithm for Learning Efficient PINN (Scale-PINN) embodies this idea: by embedding a sequential correction mechanism within the training process through an auxiliary loss function, Scale-PINN achieves both speed and sample efficiency. This perspective marks a paradigm shift in how PINN losses are conceived and constructed, by drawing on the foundations of iterative methods that have long powered scientific computing. The inherently iterative nature of these updates aligns seamlessly with the current state-of-the-art mini-batch stochastic gradient descent (SGD) optimizers. On representative stiff-conditioned lid-driven cavity benchmarks, Scale-PINN reaches a target accuracy of relative error $\leq 2e^{\text{-2}}$ in sub-2 minutes, compared with 15 hours for prior state-of-the-art training strategies~\cite{wang2025gradient}, and attains improved accuracy. A schematic diagram of Scale-PINN is illustrated in Fig.~\ref{fig:schemetic}. Our proposed framework is generic and easy to implement with broad applicability to domains such as aerodynamics and urban science, and it opens new possibilities for tackling practical problems in computational science with PINNs.

\section{Results}\label{sec:results}

% What problem the proposed method trying to solve? Complexity of PDE loss - oscillatory optimization path - slow convergence or completely stuck in local minimum. 

\subsection{Sequential correction algorithm for learning efficient PINN models} \label{sec:scale-brief}

We introduce Scale-PINN as a neural PDE solver for scientific simulations, emphasizing its capability to predict physical outcomes in fully specified systems governed by PDEs and the prescribed initial conditions (IC) and boundary conditions (BC). By reformulating the simulation as a physics-informed learning task, we seek to optimize the network weights $\boldsymbol{w}$ such that the output function $f$ satisfies the requisite PDE constraints.

The objective (loss) function for PINN weight parameters optimization can be expressed as $\mathcal{L}(\boldsymbol{w}) = \mathcal{L}_{pde} + \lambda_{ic}\ \mathcal{L}_{ic} + \lambda_{bc}\ \mathcal{L}_{bc}$ which comprises contributions from the PDE, ICs, and BCs (see Method~\ref{sec:PINN-problem}). Among the three loss components, the governing PDE loss,
\begin{align}
    \label{eq:pde_loss}
    \mathcal L_{pde} = \lVert \mathcal{N}_\vartheta[f(\cdot;\boldsymbol{w})] - h(\cdot) \rVert _{L^2(\Omega \times (0,T])}^2
\end{align}
greatly affect the PINN training difficulty. Highly nonlinear PDEs, such as the Navier-Stokes (N-S) equations, often exhibit steep gradients and strong interactions among variables. This makes it difficult for PINNs to satisfy all governing equations simultaneously. Moreover, small training perturbations can substantially change the PDE dynamics across the domain, hindering convergence to the correct solution. These factors, in addition to the high dimensionality and non-linear characteristics of neural networks, contribute to a rugged PDE loss landscape, resulting in multiple local minima, oscillatory optimization paths, and hence a higher likelihood of becoming trapped in suboptimal solutions~\cite{wong2026evolutionary}.

PINN training is commonly performed using iterative optimization methods, such as SGD and Adam algorithms. They progressively evolve the weight parameters from an initial guess $\boldsymbol{w}^{0}$, over many iterations along the descent direction of the loss function
\begin{align}
    \label{eq:SGD}
    \boldsymbol{w}^{k+1} = \boldsymbol{w}^{k} - \eta \nabla \mathcal{L}(\boldsymbol{w}^{k})
\end{align}
where the current iteration number $k$ is denoted in superscript, and $\eta$ is a problem-dependent learning rate. This approach, often implemented via mini-batch training, exploits a varying set of sample points for loss (and gradient) evaluation at each iteration. In the context of PINNs, training with incomplete and continuously changing system information introduces an additional layer of instability, amplifying optimization oscillations and making stable convergence in a complex loss landscape more difficult.

Mathematically derived from the iterative scheme (see Method~\ref{sec:scale-iterative} for details), we propose Scale-PINN by introducing a sequential correction term (auxiliary sequence) $\mathcal{F}$ at iteration $k >0$, which modifies the PDE loss term $\mathcal L_{pde}$ to improve convergence:
\begin{subequations}
    \label{eq:scpde_loss}
    \begin{align}
            &\mathcal L_\textit{sc-pde}^k = \Vert \mathcal{N}_\vartheta[f(\cdot;\boldsymbol{w}^k)] - h(\cdot)+ \frac{1}{\tau_{sc}}\mathcal{F} \Vert _{L^2(\Omega \times (0,T])}^2 \\
        &\mathcal{F} = \mathbb{B}~(f(\cdot;\boldsymbol{w}^{k})-f(\cdot;\boldsymbol{w}^{k-1}))
    \end{align}
\end{subequations}
${\tau_{sc}}$ is the hyperparameter. The matrix $\mathbb{B}$ constitutes a key design element of the iterative framework, as it determines the operator used in the update. $\mathbb{B}$ can be flexibly selected to reflect problem-dependent structure or to promote specific solution properties. The standard PINN loss function is obtained as the limiting case $\mathbb{B} = 0$.

In present study, we instantiate $\mathbb{B}\equiv\mathcal{P}_{\alpha}= (I - {\alpha^2} \nabla^2 )$ as the residual smoothing operator applied to the change in solution $f(\cdot;\boldsymbol{w}^{k})-f(\cdot;\boldsymbol{w}^{k-1})$ during iterative optimization. We show equivalence to the implicit residual smoothing method (see Method~\ref{sec:scale-full} for details), with associated enhanced stability and reduced oscillation during training. $\mathcal L_\textit{sc-pde}$ from equation (\ref{eq:scpde_loss}) can then be recast as:
\begin{subequations}
    \label{eq:scpde_loss_recast}
    \begin{align}
        \mathcal L_\textit{sc-pde}^k &= \Vert \mathcal{N}_\vartheta[f(\cdot;\boldsymbol{w}^k)] - h(\cdot)+ \frac{1}{\tau_{sc}}\mathcal{F} -\frac{\gamma}{\tau_{\alpha}}\nabla^2\mathcal{F} \Vert _{L^2(\Omega \times (0,T])}^2 \nonumber\\    
        &= \Vert \mathcal{N}_\vartheta[f(\cdot;\boldsymbol{w}^k)] - h(\cdot) + (\mathcal M_f - \mathcal M_v) \Vert _{L^2(\Omega \times (0,T])}^2  \\
        \mathcal M_f &= \frac{1}{\tau_{sc}}f(\cdot;\boldsymbol{w}^k) -\frac{\gamma}{\tau_{\alpha}}\nabla^2f(\cdot;\boldsymbol{w}^k) \\
        \mathcal M_v &= \frac{1}{\tau_{sc}}f(\cdot;\boldsymbol{w}^{k-1}) -\frac{\gamma}{\tau_{\alpha}}\nabla^2f(\cdot;\boldsymbol{w}^{k-1})
    \end{align}
\end{subequations}
with tunable hyperparameters $\tau_{sc} > 0$, $\gamma > 0$, and $\tau_{\alpha} > 0$ ($\alpha^2 = \tau_{sc} \frac{\gamma}{\tau_{\alpha}}$).

Different from standard PDE loss, two additional auxiliary terms, i.e., stabilization term $\mathcal M_f$ (residual smoothing operator) and consistency term $\mathcal M_v$ (counter term compensates for the inclusion of $\mathcal M_f$), are introduced to enhance the PINN training behavior as well as ensure the final solution will converge to original system. Equation (\ref{eq:scpde_loss_recast}) is straightforward to implement, as the auxiliary terms in $\mathcal M_f$ are already computed for the standard PDE loss. The new required operations are storing of the network weights from the previous iteration, $\boldsymbol{w}^{k-1}$, performing a forward pass to compute $f(\cdot;\boldsymbol{w}^{k-1})$, and conducting two backward passes to evaluate $\nabla^2f(\cdot;\boldsymbol{w}^{k-1})$ on the latest iteration mini-batch samples, all of which incur negligible additional computational overhead during training. Algorithm~\ref{algo:scale-pinn} (Method~\ref{sec:scale-full}) summarizes the overall computational procedure of the Scale-PINN, which integrates seamlessly with widely used iterative optimization methods such as SGD and Adam.

\begin{figure*}[h]
\centering
\includegraphics[width=1.\textwidth]{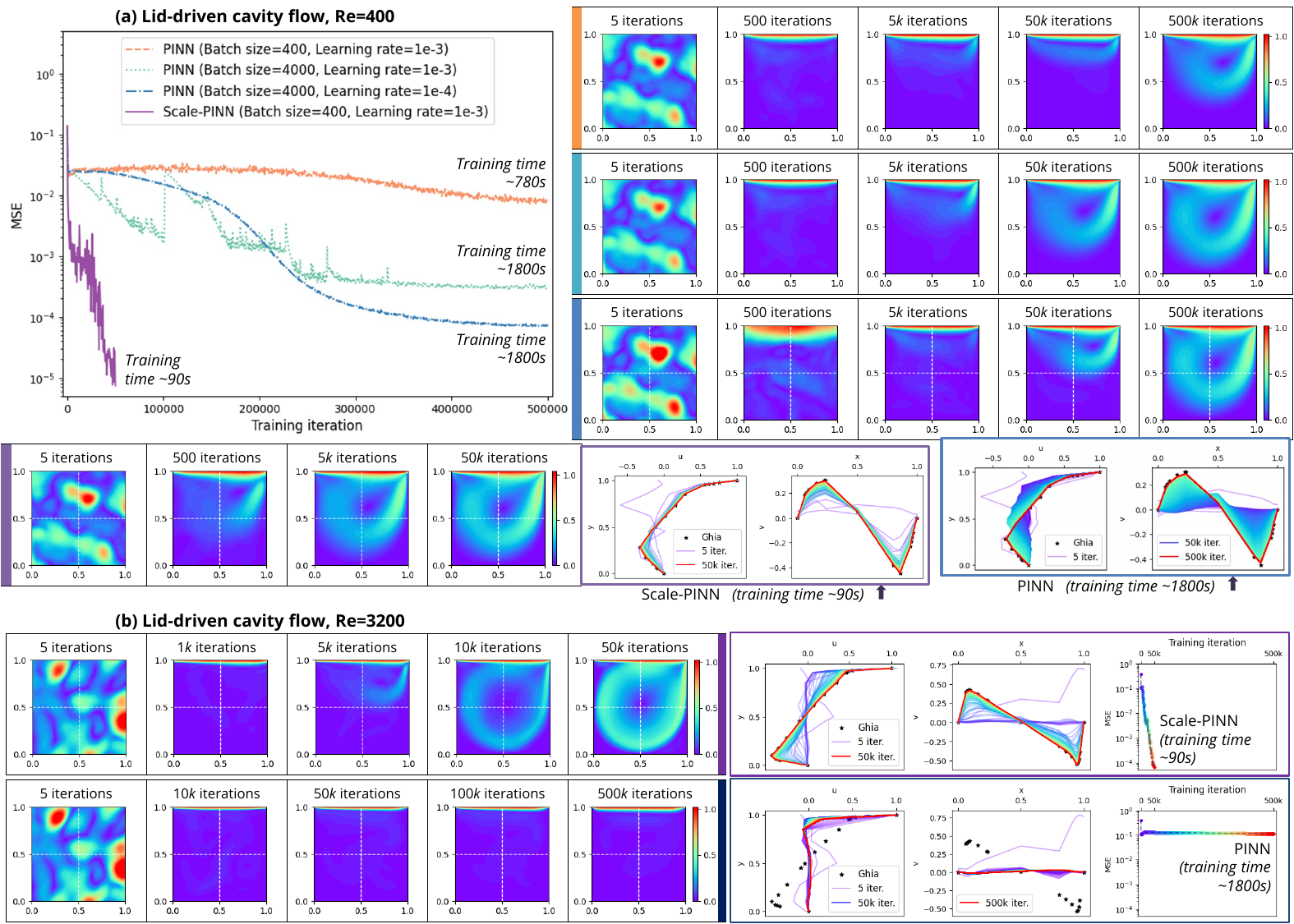}
\vspace*{-3mm}
\caption{(a) Experimental analysis on lid‑driven cavity flow at $Re=400$ shows that the convergence of a vanilla PINN can be improved by increasing batch size (400$\rightarrow$4,000) and reducing learning rate (1e$^{\text{-3}}$$\rightarrow$1e$^{\text{-4}}$), albeit at a slower pace ($\sim$1800s). Scale-PINN requires substantially less training iterations to reach orders of magnitude higher accuracy ($\sim$90s), while using 1 order of magnitude smaller batch size and higher learning rate. Comparing their intermediate flow fields progressing from a few iterations to 50$k$-500$k$ iterations, and mid‑section profiles against the Ghia \textit{et al.}~\cite{bib:Ghia82} benchmark, Scale‑PINN attains accurate flow structures far earlier. (b) Scale-PINN can converge to an accurate solution even when the Reynolds number is increased to $Re=3200$, without the need to increase batch size and number of training iterations. A vanilla PINN struggles to solve the $Re=3200$ case, as it becomes trapped in incorrect flow patterns, indicating premature convergence.}
\label{fig:ldc-1}
\end{figure*}

\subsection{Efficient scientific simulation with Scale-PINN} \label{sec:scientific-simulation}

\begin{figure*}[!ht]
\centering
\includegraphics[width=1.\textwidth]{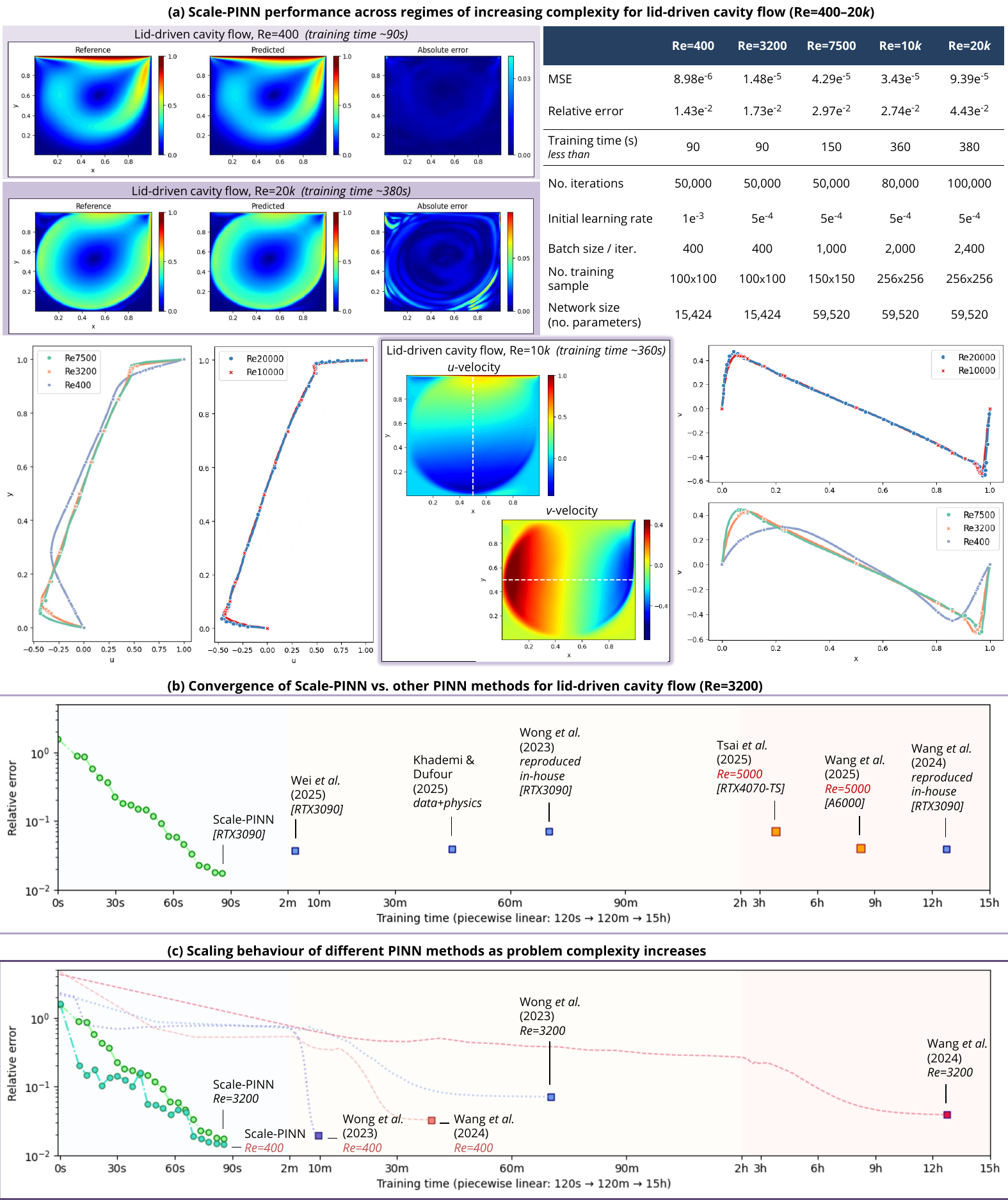}
\vspace*{-3mm}
\caption{(a) Scale‑PINN solves the lid-driven cavity flow from $Re=400$ to $Re=20k$ with state‑of‑the‑art accuracy and efficient training, as shown in the summary table of error, training time, and other parameters, alongside the representative velocity fields and absolute error maps ($Re=400$ and $Re=20k$) to confirm that residuals remain small and largely confined to shear layers and vortex cores. For all simulated cases, their MSE consistently below 1e$^{\text{-4}}$, and their mid‑section $u$ and $v$-velocity profiles (colored lines) show excellent agreement with the classic benchmark results (marked points) for numerical solvers, i.e., Ghia~\cite{bib:Ghia82} for up to $Re=10k$ and Erturk~\cite{erturk2009discussions} for $Re=20k$. (b) Scale-PINN establishes a sub-2 minutes training regime on lid-driven cavity flow ($Re=3200$), whereas recent PINN variants require hours to approach comparable accuracy. (c) Scale-PINN scales favorably with problem complexity ($Re$: 400$\rightarrow$3200), enabling the solution of more complex problems within a feasible time scale.}
\label{fig:ldc-2}
\end{figure*}

We demonstrate Scale-PINN on a classical benchmark problem in computational fluid dynamics (CFD), the lid-driven cavity flow (Method~\ref{sec:ns-equation}). The fluid flow inside a 2D unit square, $x \in [0,1]$, $y \in [0,1]$, is driven by the top lid velocity ($u_{lid} = 1$), and governed by the steady-state incompressible N-S equations for velocity $\vec{u}=[u,v]^\intercal$ and pressure $p$:
\begin{subequations}
    \label{eq:NS}
    \begin{align}
        \label{eq:continuity}
        \nabla\cdot \vec{u} &= 0 \\
        \label{eq:momentum}
        (\vec{u}\cdot\nabla)\vec{u} &= \frac{1}{Re}\nabla^2 \vec{u} - \nabla p
    \end{align}
\end{subequations}
Complex physical phenomenon can be observed when the Reynolds number ($Re$) increases, such as $Re \geq$ 3200, making it notoriously difficult for PINN methods to solve (e.g., require hours to tens of hours of training) even with the help of some labeled data or transfer and curriculum learning~\cite{wang2023experts, wang2024piratenets}. In contrast, Scale-PINN is fast and effective at tackling this very challenging PINN benchmark problem (see results highlighted in Fig.~\ref{fig:schemetic}-Fig.~\ref{fig:ldc-2}).

Building on equation (\ref{eq:scpde_loss_recast}) as per the Scale-PINN methodology, the loss function for momentum equations (\ref{eq:momentum}) can be defined as below:
\begin{subequations}
\label{eq:momentum-scale}
\small
\begin{align}
        \mathcal L_\textit{sc-pde(Mu)}^k &= \Vert u^k\frac{\partial u^k}{\partial x} + v^k\frac{\partial u^k}{\partial y} -\frac{1}{Re}(\frac{\partial^2 u^k}{\partial x^2} + \frac{\partial^2 u^k}{\partial y^2}) + \frac{\partial p^k}{\partial x} + \mathcal S_\textit{Mu} \Vert_{L^2(\Omega)}^2 \\
        \mathcal L_\textit{sc-pde(Mv)}^k &= \Vert u^k\frac{\partial v^k}{\partial x} + v^k\frac{\partial v^k}{\partial y} -\frac{1}{Re}(\frac{\partial^2 v^k}{\partial x^2} + \frac{\partial^2 v^k}{\partial y^2}) + \frac{\partial p^k}{\partial y} + \mathcal S_\textit{Mv} \Vert_{L^2(\Omega )}^2 \\        
        \mathcal S_\textit{Mu} &= \frac{1}{\tau_{sc}}(u^k-u^{k-1}) -\frac{\gamma_{uv}}{\tau_{\alpha}}\left[ (\frac{\partial^2 u^k}{\partial x^2} + \frac{\partial^2 u^k}{\partial y^2})-(\frac{\partial^2 u^{k-1}}{\partial x^2} + \frac{\partial^2 u^{k-1}}{\partial y^2}) \right] \\
        \mathcal S_\textit{Mv} &= \frac{1}{\tau_{sc}}(v^k-v^{k-1}) -\frac{\gamma_{uv}}{\tau_{\alpha}}\left[ (\frac{\partial^2 v^k}{\partial x^2} + \frac{\partial^2 v^k}{\partial y^2})-(\frac{\partial^2 v^{k-1}}{\partial x^2} + \frac{\partial^2 v^{k-1}}{\partial y^2}) \right]
\end{align}
\end{subequations}
with the integration of sequential correction terms $S_\textit{Mu}$ and $\mathcal S_\textit{Mv}$ as defined by the choice of the Helmholtz residual smoothing operator ($\mathcal{P}_{\alpha} = (I - {\alpha^2} \nabla^2 )$). In this study we set $\gamma_{uv} = \frac{1}{Re}$ based on prior knowledge of the governing physical system. Values of $\tau_{sc}$ and $\tau_{\alpha}$ are fine-tuned empirically.

A fundamental difficulty in solving incompressible N-S equations is that the pressure variable does not appear explicitly in the continuity equation and only appears through its gradient in the momentum equations~\cite{Toutant17,Toutant18}. Thus, several numerical schemes proposed to modify the continuity formulation to relax the incompressibility constraint by explicitly establishing a dynamic relationship between pressure and continuity ~\cite{bib:Chorin67,Toutant18,bib:Chiu18}. Guided by the same principle, the Scale-PINN loss function for the continuity equation (\ref{eq:continuity}) is defined as follows:
\begin{subequations}
\label{eq:continuity-scale}
\begin{align}
        \mathcal L_\textit{sc-pde(Cn)}^k &= \Vert \frac{\partial u^k}{\partial x} + \frac{\partial v^k}{\partial y} + \mathcal S_\textit{Cn} \Vert_{L^2(\Omega)}^2 \\  
        \mathcal S_\textit{Cn} &= \frac{1}{\tau_{sc}}(p^k-p^{k-1})
\end{align}
\end{subequations}
Scale-PINN allows $\mathcal S_\textit{Cn}$ to be introduced into the continuity loss term to explicitly provide a relation between pressure and the continuity equation, thereby improving convergence. Together with a BC loss term $\mathcal{L}_{bc}$ to enforce the top lid velocity $u_{lid} = 1$, $v_{lid} = 0$ and no-slip wall condition $u = v = 0$, the Scale-PINN objective function for simulating the lid-driven cavity flow is thus defined as: $\mathcal{L}_\textit{sc}(\boldsymbol{w}^k) = \mathcal L_\textit{sc-pde(Mu)}^k +  \mathcal L_\textit{sc-pde(Mv)}^k + \mathcal L_\textit{sc-pde(Cn)}^k + \lambda_{bc} \mathcal{L}_{bc}^k$.

Our experimental analysis (Fig.~\ref{fig:ldc-1}) suggests that the principal barrier for vanilla PINNs is not merely insufficient compute but an unstable optimization landscape. At $Re=400$, with a small batch size and large learning rate, the convergence is slow (over 500$k$ iterations, training time $\sim$780s) and susceptible to premature locking into suboptimal flow patterns. Increasing the batch ten-fold (400$\rightarrow$4,000) and reducing the learning rate ten-fold (1e$^{\text{-3}}$$\rightarrow$1e$^{\text{-4}}$) stabilizes convergence, but it also leads to significantly increased training time ($\sim$1800s). In contrast, Scale-PINN attains much lower error  (MSE $<$ 1e$^{\text{-5}}$) in only 50$k$ iterations (training time $\sim$90s) using the smaller batch and larger learning rate. Visual snapshot across training iterations show that Scale-PINN resolves the primary vortex and corner eddies earlier, and its mid‑section $u‑$ and $v‑$profiles match the classical Ghia \textit{et al.}~\cite{bib:Ghia82} cut‑lines, indicating physically faithful pressure-velocity coupling. At $Re=3200$, Scale-PINN continues to converge without increasing batch size or iteration budget, whereas vanilla PINN becomes trapped in incorrect local minima. These contrast supports our interpretation that the sequential correction in Scale-PINN effectively smooths the PDE residuals, enabling steady progress with standard first-order optimizers.

The cross‑regime results summarized in Fig.~\ref{fig:ldc-2}(a) underscore that Scale-PINN not only attains state-of-the-art accuracy but also establishes a new benchmark in efficiency across an unprecedented range of Reynolds numbers. Scale-PINN remains both accurate and fast from $Re=400$ to $Re=20k$, with relative error rising only modestly (1.4e$^{\text{-2}}$$\rightarrow$4.4e$^{\text{-2}}$) as the problem complexity increases (learning stiffness intensifies), while training time stays under seven minutes ($\sim$380s) even at $Re=20k$. Scale-PINN demonstrates favorable scaling of optimization cost with problem complexity, despite deliberately increased resolution (100$^{\text{2}}$$\rightarrow$256$^{\text{2}}$), batch size (400$\rightarrow$2400), and number of iterations (50$k$$\rightarrow$100$k$) to resolve thinner boundary layers and stronger shear at higher Reynolds number. For all simulated cases, their mid‑section $u$ and $v$-velocity profiles show excellent agreement with canonical benchmarks (Ghia \textit{et al.}~\cite{bib:Ghia82} up to $Re=10k$ and Erturk~\cite{erturk2009discussions} at $Re=20k$), thereby demonstrating, for the first time, that a PINN approach can deliver accurate and efficient N-S solutions at the high-Reynolds regime. The results highlight Scale-PINN as a scalable method across regimes, with predictions consistent with established numerical standards.

We benchmark Scale-PINN against recent PINN methods at $Re=3200$, which features in several state-of-the-art PINN literature as a challenging regime, where many PINNs fail to converge. We include methods whose original paper reports successfully solving the problem with relative error below 1e$^{\text{-1}}$: Wong \textit{et al.} (2023, LSA-PINN)~\cite{wong2023lsa}, Wang \textit{et al.} (2024, PirateNets)~\cite{wang2024piratenets}, Khademi \& Dufour (2025, TSA-PINN)~\cite{khademi2025physics}, and Wei \textit{et al.} (2025, FFV-PINN)~\cite{wei2025ffv}. LSA-PINN and PirateNets are reproduced in-house on a single RTX 3090 (Scale-PINN trained on the same hardware) using available source code. We also include $Re=5000$ results from Wang \textit{et al.} (2025, SOAP)~\cite{wang2025gradient}, a second-order optimizer and successor to PirateNets, and Tsai \textit{et al.} (2025, MLD-PINN)~\cite{tsai2025mld}, because they disclose timing; this offers a useful speed-accuracy reference despite the higher Reynolds number. As shown in Fig.~\ref{fig:ldc-2}(b), Scale-PINN's sub-2 minutes training regime represents a new state-of-the-art speed-accuracy Pareto frontier for $Re=3200$. Scale-PINN is trained from scratch, i.e., He initialization~\cite{he2015delving}, without pre-training, curriculum schedules, or data supervision. Contemporary PINN variants are often aided by curriculum strategies (e.g., $Re$: 100$\rightarrow$400$\rightarrow$1000$\rightarrow$...$\rightarrow$3200)~\cite{wang2024piratenets, wang2025gradient}, numerical differentiation loss~\cite{wong2023lsa, wei2025ffv, tsai2025mld}, or additional data supervision~\cite{khademi2025physics}, and many of them still require hours of training to approach comparable prediction accuracy.

Fig.~\ref{fig:ldc-2}(c) illustrates the convergence behavior of Scale-PINN relative to other methods as the problem complexity increases ($Re$: 400$\rightarrow$3200). Scale-PINN maintains excellent convergence speed and accuracy across regimes. In contrast, LSA-PINN training increases from less than 10 minutes to more than 1 hour, accompanied by a significant degradation in accuracy, while PirateNets requires up to 12-15 hours to achieve comparable accuracy. This contrast highlights the superior scalability of Scale-PINN: as the problem becomes more challenging, its convergence speed degrades far more slowly than competing methods, enabling the solution of increasingly complex fluid dynamics problems within feasible computational budgets.

The impressive improvements in Scale-PINN training regime motivate, for the first time in a PINN study, direct comparisons with high-fidelity CFD solvers. We compare Scale-PINN against an in-house CFD solver~\cite{bib:Chiu21}, which has been demonstrated to produce high-fidelity solutions for the incompressible N-S equations, and the widely-used commercial solver Ansys Fluent, recognized for its reliability, generality, and parallel performance.% Mesh-independence tests confirm that a $512 \times 512$ mesh provides a converged ground-truth solution for the lid-driven cavity flow at $Re=3200$.

We first run the in-house solver on a $96\times 96$ mesh and record the runtime and accuracy along the convergence trajectory. Given a runtime of around 80s, Scale-PINN achieves higher converged accuracy (Fig.~\ref{fig:schemetic}), establishing a new speed-accuracy Pareto frontier. We then utilize similar simulation settings in Fluent on a finer $128 \times 128$ mesh. Its converged accuracy is much poorer than both the in-house code (on $96\times 96$ mesh) and Scale-PINN. Finally, we perform the simulations on a much finer $192 \times 192$ mesh using four CPU cores for both Fluent and the in-house code. With the optimized parallelization, results from Fluent now match our method's accuracy in around 120s runtime. The in-house solver, with a basic OpenMP implementation, took nearly twice as long to achieve the same accuracy, although it should be acknowledged that it can eventually reach the highest accuracy with additional runtime.

These comparisons suggest that mesh-free PINN methods may be well suited for practical scientific and engineering problems under fixed computational budgets. In scientific computing, practitioners often choose the coarsest mesh that delivers physics-resolved accuracy, balancing fidelity against cost; achieving higher accuracy typically requires finer meshes and substantially more compute. Leveraging theoretically exact automatic differentiation for derivative evaluation, Fig.~\ref{fig:schemetic} shows that Scale-PINN can outperform a conventional second-order numerical scheme as employed by Fluent on a relatively coarse mesh (e.g., $128 \times 128)$. In practical terms, under a limited time budget, Fluent on a $128 \times 128$ mesh yields the best accuracy within $\sim$30s runtime; Scale-PINN provides the best accuracy within $\sim$120s; and if runtime is unconstrained, the in-house solver on a fine mesh achieves the highest accuracy.

\subsection{Navier–Stokes flow simulation for engineering and urban science} \label{sec:application-results}

\begin{figure*}[!ht]
\centering
\includegraphics[width=1.\textwidth]{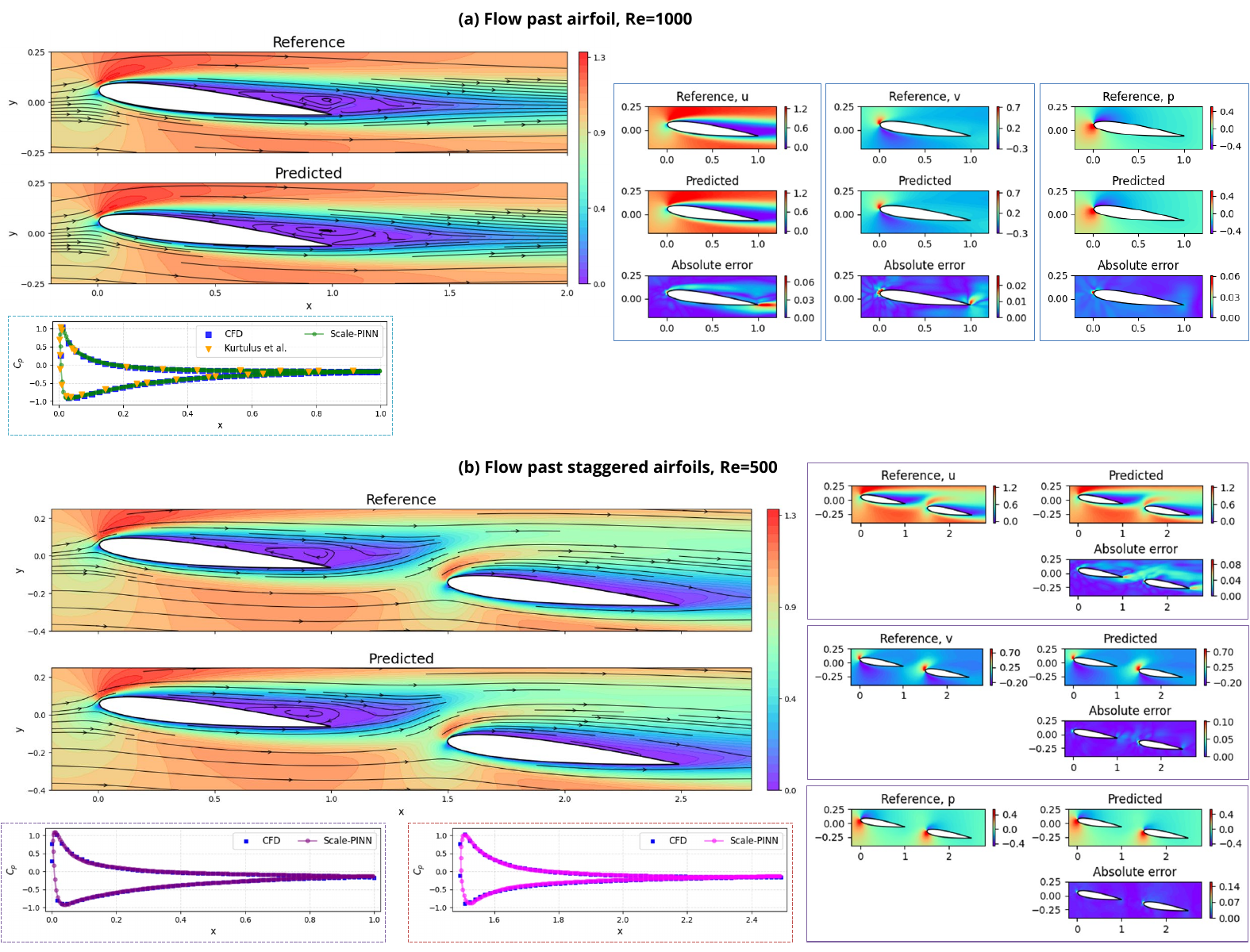}
\vspace*{-3mm}
\caption{Scale-PINN predictions, with streamlines overlaid on velocity magnitude contours, for flow past (a) a single-airfoil at $Re=1000$ and (b) staggered airfoils at $Re=500$ are compared to reference solutions obtained from CFD. Component-wise fields ($u$, $v$, $p$) and absolute error maps indicate good agreement between Scale-PINN and CFD across the domain, including wakes behind airfoil. Surface pressure‑coefficient ($C_{p}$) traces along the airfoil(s) closely match CFD and literature curves. Only the near-field region is shown for clarity, where the flow patterns around the airfoil(s) emerge; the actual computational domain extends well beyond the visualized region. Scale-PINN reaches accurate solutions within $\sim$180s of training, achieving near-field velocity relative errors of $1.7e^{\text{-2}}$ ($3.47e^{\text{-3}}$ full domain) for the single-airfoil case and $1.96e^{\text{-2}}$ ($4.79e^{\text{-3}}$ full domain) for the staggered airfoils case. These results validate the accuracy and efficiency of Scale-PINN in resolving canonical aerodynamic flow features at moderate Reynolds number.}
\label{fig:flow-past-foil}
\end{figure*}

To demonstrate the applicability beyond lid-driven cavity benchmarks, we present Scale-PINN results on two representative aerodynamic problems: (1) flow past a single NACA0012 airfoil at $Re=1000$ and $7^\circ$ angle of attack (AoA); and (2) flow past two-staggered NACA0012 airfoils at $Re=500$ and $7^\circ$ AoA (Method~\ref{sec:ns-equation}). NACA airfoil problems are classical benchmarks, where fast and reliable flow simulations are crucial for advancing the design and optimization of aerodynamic structures such as aircraft and wind turbines.

For the single-airfoil case, Scale-PINN predictions produce the expected wake patterns (Fig.~\ref{fig:flow-past-foil}a) and show excellent agreement between component-wise velocity ($u$, $v$) and pressure ($p$) fields and CFD reference solutions. Even for the more complex scenario of two-staggered airfoils, Scale-PINN successfully captures the altered wake structures and aerodynamic interactions between the bodies, with predicted velocity and pressure fields aligning well with CFD (Fig.~\ref{fig:flow-past-foil}b). For both cases, absolute error maps indicate that minor and localized discrepancies are mainly confined within the leading edge and near-wake regions. The pressure coefficient ($C_{p}$) distribution along the airfoil surfaces compare well with both CFD and literature data from Kurtulus \cite{bib:Kurtulus2015}, even capturing the suction peak at the leading edge and subsequent pressure recovery.

Scale-PINN requires only $\sim$180s of training to produce solutions in excellent agreement with CFD ground truth. In contrast, Xiao \textit{et al.}~\cite{bib:Xiao25} report a training time of $\sim$7800s and the need for additional sensor points to achieve comparable accuracy for airfoil flows, highlighting the superior efficiency and accuracy of Scale-PINN. Scale-PINN demonstrates robust performance across both single-body configurations and multi-body aerodynamic interactions, underscoring its potential for aerodynamic design and optimization with multi-element airfoils.

\begin{figure*}[h]
\centering
\includegraphics[width=1.\textwidth]{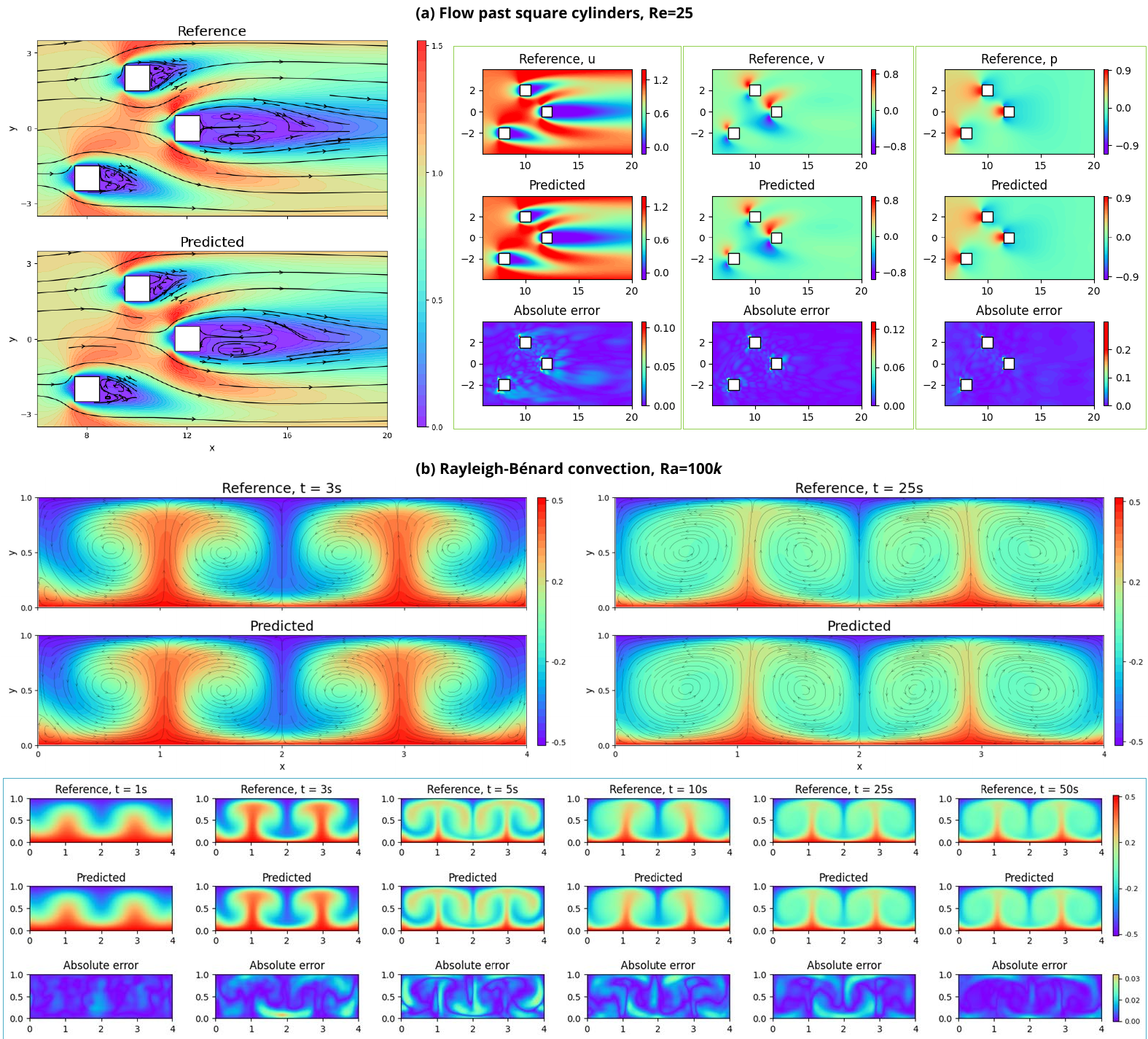}
\vspace*{-3mm}
\caption{(a) Scale-PINN simulates the flow past square cylinders in open domain, where the predicted fields ($u$, $v$, $p$) accurately capture the wake structure and recirculation zones, with consistently low absolute errors against the reference solution obtained from CFD. Contours are shown in the near field for clarity; the actual computational domain extends further. Scale-PINN reaches accurate solutions within $\sim$285s of training, achieving near-field velocity relative errors of $9.21e^{\text{-3}}$ ($4.86e^{\text{-3}}$ full domain). (b) Scale-PINN simulates Rayleigh–Bénard convection at $Ra=100k$. The temperature contours, overlaid with velocity streamlines, show close agreement between predicted roll patterns and reference solutions obtained from CFD, across multiple time snapshots (1-50s). Scale-PINN achieves accurate solutions within $\sim$390s of training, with relative errors $1.99e^{\text{-2}}$ for velocity and $3.2e^{\text{-2}}$ for temperature. These cases validate the method’s robustness across bluff-body geometries in open domain and thermally driven, time-dependent convection.}
\label{fig:urban}
\end{figure*}

We next simulate flow past three staggered square cylinders at $Re=25$, a canonical proxy for wind flow around buildings (Method~\ref{sec:ns-equation}). Scale-PINN accurately recovers wake structures and recirculation zones, with velocity and pressure fields closely matching CFD references and errors confined to shear layers and separation points (Fig.~\ref{fig:urban}a). These results confirm Scale-PINN’s robustness for bluff-body aerodynamics in open domains, where reliable prediction of flow interactions are commonly used to inform urban ventilation design and pollutant dispersion.

We further validate Scale-PINN on buoyancy-driven Rayleigh–Bénard convection at $Ra=100k$ to demonstrate its versatility in modeling multiphysics transient dynamics, where thermal instabilities give rise to convection rolls and their transient evolution into complex patterns (Method~\ref{sec:ns-equation}). Scale-PINN predictions remain in close agreement with CFD benchmarks across multiple time snapshots between 1s and 50s, with low residuals sustained over time (Fig.~\ref{fig:urban}b). By capturing the onset and development of natural convection, Scale-PINN demonstrates its capacity to model thermally-driven flows of direct importance to energy efficiency, ventilation, and thermal comfort in the built environment and urban sustainability studies.

Our results further confirm the superior sampling points (or mesh fidelity) to accuracy trade-off enabled by Scale-PINN (relative to CFD). With temporal step size of 0.001, CFD solution achieves an MSE of $4.0e^{\text{-5}}$ and $3.32e^{\text{-5}}$ for temperature and velocity magnitude, respectively, on a fine spatial mesh ($384\times96$). It only achieves an MSE of $1.3e^{\text{-4}}$ and $1.1e^{\text{-4}}$ with a coarse mesh ($256\times64$). Scale-PINN with $258\times 66\times 501$ spatio-temporal sample points, which is on par with CFD coarse mesh, can produce temperature and velocity MSEs' of 4.5e$^{\text{-5}}$ and 3.1e$^{\text{-5}}$, respectively. This further highlights the potential for PINNs to have a unique place within the Pareto set of models for use when one might need to trade-off accuracy and computational cost (time) for scientific simulations.

\begin{figure*}[!ht]
\centering
\includegraphics[width=1.\textwidth]{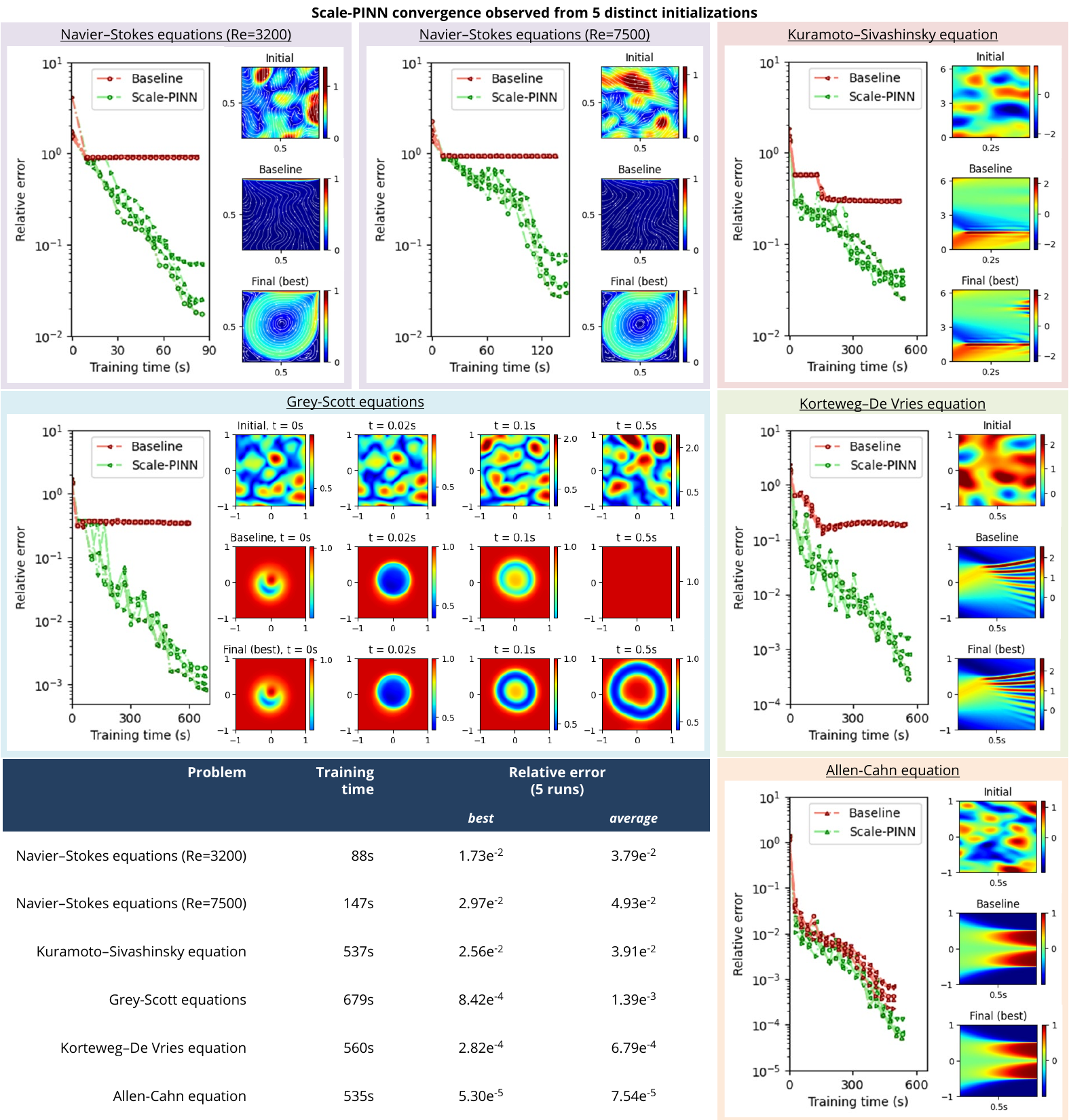}
\vspace*{-3mm}
\caption{Scale-PINN attains state-of-the-art, minute‑scale training efficiency on a range of PDE benchmarks, converging within $\sim$10 minutes while achieving accuracy on par with second-order optimization method (typically requires hours of training). For each PDE benchmark problem, relative error (log scale) versus training time is shown for five independent initializations; the \textbf{final (best)} Scale-PINN solution, final \textbf{baseline} PINN solution, and their corresponding \textbf{initial} (random initialization) solution are shown alongside error-time curves. We keep the model architecture (moderate-sized multilayer perceptron) and training settings identical and change only the PDE residual term, i.e., sequential corrected (Scale-PINN) vs. standard PDE loss (baseline). The accompanying table reports training times and best/average errors over the five runs, demonstrating robustness to initialization and broad applicability across PDE families.}
\label{fig:ablation}
\end{figure*}

\subsection{Performance on benchmark problems} \label{sec:benchmark-results}

We demonstrate the effectiveness of Scale-PINN across several benchmark PDE problems: Kuramoto-Sivashinsky, Grey-Scott, Korteweg-de Vries, Allen-Cahn, and N-S equations at $Re=3200, 7500$. Detailed descriptions of the problems are given in Method~\ref{sec:PINN-simulation}. These PDEs span a wide range of physical phenomena and application domains: N-S underpins fluid dynamics; Kuramoto–Sivashinsky models instabilities in physical systems such as chemical reaction dynamics and thin-film flows; Grey-Scott captures reaction-diffusion pattern formation in chemistry and biology; Korteweg-de Vries describes shallow-water and plasma solitary waves; and Allen-Cahn governs phase separation and interface motion in materials science. To isolate the contribution of our sequential corrected loss from conventional PDE loss in PINN training, we perform a simple ablation: we keep the model architecture (moderate-sized multilayer perceptron) and training settings identical and change only the PDE loss term, i.e., sequential corrected (Scale-PINN) vs. standard PDE loss (baseline). Each problem is run for five optimization trails from distinct model initializations using He method~\cite{he2015delving}, evaluating robustness to initialization. Their error-time convergence and representative initial and final solutions are illustrated in Fig.~\ref{fig:ablation}, showing rapid convergence of Scale-PINN. A concise summary of accuracy and runtime is also provided.

Scale-PINN substantially accelerates learning to achieve accurate solutions within 10 minutes, whereas the baseline fails to simulate the correct patterns for N-S, Kuramoto-Sivashinsky, and Grey-Scott equations. The ablation validates that the sequential corrected PDE loss is the driver of the gains. Notably, our prediction accuracy on the benchmark PDE problems approaches that reported by Wang \textit{et al.} (2025, SOAP)~\cite{wang2025gradient} which uses second-order optimizer for PINN and requires hours to tens of hours of training.

\section{Discussion} \label{sec:discussion}

Scale-PINN reframes how physics-informed loss functions are conceived and constructed: moving beyond PDE and discretization towards embedding the principle of iterative methods directly into the loss formulation. By introducing a sequential residual-correction mechanism, Scale-PINN converges rapidly and stably without sacrificing accuracy, establishing it as a state-of-the-art neural PDE solver. More broadly, this reformulation encourages the computational science community to regard the loss function not merely as an error metric, but as a mechanism for encoding the mathematics of convergence.

The framework is designed for immediate adoption and re-engineering. In this work, the residual-smoothing operator is chosen and shown to be advantageous across diverse physical systems, and can be readily adapted to different architectures and optimizers. Nonetheless, the sequential-correction principle naturally extends to incorporation of algorithmic insights from other iterative methods in scientific computing (described in Methods~\ref{sec:scale-iterative}~\&~\ref{sec:scale-full}). Beyond its algorithmic contribution, our work offers a conceptual bridge between scientific computing and modern AI. We anticipate that this sequential correction learning paradigm will stimulate a new generation of physics-informed learning frameworks that fuse scientific computing and machine learning more coherently, advancing PINNs toward the reliability, scalability, and rigor long achieved by traditional numerical methods.

%Conclusions may be used to restate your hypothesis or research question, restate your major findings, explain the relevance and the added value of your work, highlight any limitations of your study, describe future directions for research and recommendations. 

%In some disciplines use of Discussion or 'Conclusion' is interchangeable. It is not mandatory to use both. Please refer to Journal-level guidance for any specific requirements. 

% Don't do words count from now on!
%TC:ignore

\section{Methods} \label{sec:method}

\subsection{PINN models for scientific simulations} \label{sec:PINN-problem}

PINNs are a class of universal function approximators capable of learning a mapping $f$ between the input variables $(x,t)$ and output solution $u$ while satisfying specified differential equation constraints that represent the physical phenomenon or dynamical process of interest. Consider differential equations of the general form:
\begin{subequations} \label{eq:pde_ibc_eqn}
    \begin{align}
        & \text{PDE:} & \mathcal{N}_\vartheta[u(x,t)] &= h(x,t), & x\in\Omega, t\in(0,T] \label{eq:pde_eqn} \\
        & \text{IC:} & u(x,t=0) &= u_0(x), & x\in\Omega \label{eq:ic_eqn} \\
        & \text{BC:} & \mathcal{B}[u(x,t)] &= g(x,t), & x\in\partial\Omega, t\in(0,T] \label{eq:bc_eqn}
    \end{align}
\end{subequations} 
where the general differential operator $\mathcal{N}_\vartheta[u(x,t)]$ can be parameterized by $\vartheta$ and can include linear and/or nonlinear combinations of temporal and spatial derivatives of $u$, with an arbitrary source term $h(x,t)$, in the computational domain $x\in\Omega, t\in(0,T]$. The equation (\ref{eq:ic_eqn}) specifies the initial condition (IC), $u_0(x)$, at time $t=0$. The equation (\ref{eq:bc_eqn}) specifies the boundary condition (BC) at the domain boundary $\partial\Omega$ that $\mathcal{B}[u(x,t)]$ equates to $g(x,t)$, where $\mathcal{B}[\cdot]$ can either be an identity (Dirichlet BC), a differential (Neumann BC), or a mixed identity-differential (Robin BC) operator.

Fundamentally, a PINN model can arrive at an accurate and physics-compliant prediction by forcing its output $f(\cdot;\boldsymbol{w})$ function to satisfy equation (\ref{eq:pde_ibc_eqn}) through training, i.e., optimizing its network weight parameters $\boldsymbol{w}$. The objective (loss) function of PINN weight parameters optimization can be written as:
\begin{subequations} \label{eq:pinn_loss_fn}
    \begin{align}
    \mathcal{L}(\boldsymbol{w}) &= \mathcal{L}_{pde} + \lambda_{ic}\ \mathcal{L}_{ic} + \lambda_{bc}\ \mathcal{L}_{bc} \\
    \label{eq:Loss-pde}
    \mathcal{L}_{pde} &= \lVert \mathcal{N}_\vartheta[f(\cdot;\boldsymbol{w})] - h(\cdot) \rVert _{L^2(\Omega \times (0,T])}^2 \\
    \mathcal{L}_{ic} &= \lVert f(\cdot,t=0;\boldsymbol{w}) - u_0(\cdot) \rVert _{L^2(\Omega)}^2 \label{eq:pinn_ic_fn} \\
    \mathcal{L}_{bc} &= \lVert \mathcal{B}[f(\cdot;\boldsymbol{w})] - g(\cdot) \rVert _{L^2(\partial\Omega \times (0,T])}^2 \label{eq:pinn_bc_fn}
    \end{align}
\end{subequations} 
and these are to be evaluated on a set of completely label-free training points (collocation points) sampled from the respective spatio-temporal domain during PINN training. The PINN loss usually consists of multiple components for PDEs, ICs, and BCs, where the incorporation of relative weights $\lambda_{ic} \geq 0$ and $\lambda_{bc} \geq 0$ is essential to control the trade-off between these components.

PINN models usually have a rugged loss landscape, resulting in multiple local minima, oscillatory optimization paths, and a higher likelihood of becoming trapped in suboptimal solutions during training. Consequently, large batch sizes and small learning rates are often needed to stabilize PINN training, but they significantly increase computation time and remain vulnerable to premature convergence. Many studies also adopt curriculum learning, where the model training starts with the PDE settings of an easier problem and gradually transitions to the target problem. However, this approach requires manually setting up intermediate problems and training configurations, demands a good understanding of the system's behavior under changing PDE settings, and involves long training times through solving multiple intermediate problems.

\subsection{Integration of iterative solver principle into physics-informed learning} \label{sec:scale-iterative}

To solve equation (\ref{eq:pde_ibc_eqn}), nearly all conventional numerical simulation approaches begin by converting the continuous governing PDEs into a finite-dimensional linear system, 
\begin{align}
    \label{eq:discretization}
    \mathbb{A} u = h
\end{align}
where $\mathbb{A}$ is the coefficient matrix determined by the chosen discretization scheme, $u$ is the solution vector and $h$ is the source term vector. To efficiently solve the above linear system, iterative numerical algorithms have been continually developed and refined over decades, and this is one of the cornerstones of the field of scientific computing~\cite{bib:Saad03}. The fundamental idea is that, rather than solving the linear system directly, residual-based error corrections are introduced to incrementally improve the solution in sequence. This leads to a generic formulation:
\begin{subequations}
    \label{eq:generic-iter}
    \begin{align}
        u^{k+1} = u^k + \mathbb{B}^{-1}~r^k \\
        r^k = h - \mathbb{A}u^k
    \end{align}
\end{subequations}
where $\mathbb{B}$ is the key matrix designed to mitigate the computational cost of matrix inversion, thereby improving the robustness and efficiency of the iterative algorithm and enabling memory- and time-efficient computations~\cite{bib:Xu92Iterative,bib:Morton_Mayers_2005}.

A classical example is the modified Richardson iteration:
\begin{align}
    \label{eq:Richardson}
    u^{k+1} = u^{k} + \xi~r^k
\end{align}
where $\xi > 0$ is the relaxation factor ensuring convergence of the solution~\cite{bib:Saad03}. Another example is the Jacobi iterative method:
\begin{subequations}
    \begin{align}
        \label{eq:Jacobi-iter}
        u^{k+1} = u^k + \mathbb{D}^{-1}~r^{k} \\
        \mathbb{A} = \mathbb{D} + \mathbb{L} + \mathbb{U}
    \end{align}
\end{subequations}
In the above, $\mathbb{D}$ is the diagonal matrix, and $\mathbb{L}$ and $\mathbb{U}$ denote the lower and upper triangular parts of $\mathbb{A}$. Similarly, the Gauss–Seidel iterative method can be written as:
\begin{align}
    \label{eq:GS-iter}
    u^{k+1} = u^k + (\mathbb{D}+\mathbb{L})^{-1}~r^{k}
\end{align}

We mathematically demonstrate that this iterative residual-correction principle can be effectively integrated into physics-informed learning and explicitly realized within the PINN loss formulation. We begin by reformulating the generic iterative residual-correction in equation (\ref{eq:generic-iter}), into the following loss function form:
\begin{align}
    \label{eq:loss-generic}
    \mathcal L_\textit{sc-ND}^k = \mathbb{B}~(u^{k+1} - u^k) + (\mathbb{A}u^{k}-h)
\end{align}
for any intermediate iteration step $k > 0$. However, in practice, it is infeasible to directly employ the above expression, as $u^{k+1}$ is unknown and must be estimated from the known $u^k$. To bridge this gap, we adopt a second-order extrapolation based on the Taylor-series expansion:
\begin{align}
    u^{k+1} = 2u^{k} - u^{k-1}
\end{align}
which yields the following reformulated loss:
\begin{align}
    \label{eq:lloss-generic2}
    \mathcal L_\textit{sc-ND}^k = \mathbb{B}~(u^{k} - u^{k-1}) + (\mathbb{A}u^{k}-h)
\end{align}

In the context of PINN training, the change in the solution $u^{k} - u^{k-1}$ can be approximated by the PINN model predictions $f(\cdot;\boldsymbol{w}^{k})-f(\cdot;\boldsymbol{w}^{k-1})$, while the PDE residuals, $\mathbb{A}u^{k}-h$ are represented by $\mathcal{N}_\vartheta[f(\cdot;\boldsymbol{w})] - h(\cdot)$ during iterative optimization. Without constraining to any specific discretization scheme, the generic iterative residual-correction PINN loss can therefore be expressed as:
\begin{equation}
    \mathcal L_\textit{sc-pde}^k = \Vert \mathcal{N}_\vartheta[f(\cdot;\boldsymbol{w}^k)] - h(\cdot)+ \mathbb{B}~(f(\cdot;\boldsymbol{w}^{k})-f(\cdot;\boldsymbol{w}^{k-1}))  \Vert _{L^2(\Omega \times (0,T])}^2
\end{equation}
This formulation allows the loss to be computed flexibly via automatic differentiation at arbitrary sample locations. We refer to the term $\mathbb{B}~(f(\cdot;\boldsymbol{w}^{k})-f(\cdot;\boldsymbol{w}^{k-1}))$ as a sequential correction term (or auxiliary sequence), which augments the conventional PDE loss. This leads to what we call the sequential correction loss, a paradigm shift in how PINN losses are formulated that draws on the foundations of iterative methods that have long powered scientific computing. The design of the matrix $\mathbb{B}$ plays a key role in this iterative framework. Notably, the conventional PINN loss function is recovered as a special case when $\mathbb{B} = 0$.

\subsection{Sequential correction algorithm for learning efficient PINN models} \label{sec:scale-full}

This section presents a specific formulation of the sequential correction loss using a special residual-smoothing operator $\mathcal{P}_{\alpha}$ derived from the implicit residual smoothing method and realized through modified Richardson iteration, which is parametrized by $\alpha$ and offers several desirable theoretical properties.

For numerical approaches, it is essential to employ high-order, high-resolution numerical schemes together with a mesh of sufficient resolution to accurately capture complex physical features. However, high-order schemes and fine meshes impose severe restrictions on the time-step size, i.e., often requiring it to be very small, when the problem is solved using an explicit scheme~\cite{bib:Choi90}. This is analogous to PINN training, where stable convergence often depends on small learning rates, large batch sizes, and curriculum learning. These stability constraints make convergence slow and computationally expensive in both cases.

The implicit residual-smoothing method, rooted in the traditions of scientific computing, modifies the residuals using smoothing operators before each update step~\cite{bib:CINNELLA16, bib:Wesseling01}. It has been shown theoretically that this technique alleviates stability-related limitations in numerical simulations~\cite{bib:BIENNER24} and can significantly accelerate convergence by permitting larger time steps.

Scale-PINN aims to improve convergence behavior by instantiating the sequential correction term through a residual-smoothing operator $\mathcal{P}_{\alpha}\left[ f(\cdot;\boldsymbol{w}^{k})-f(\cdot;\boldsymbol{w}^{k-1}) \right]$, thereby achieving more efficient PINN training. The proposed sequential correction algorithm can then be seamlessly integrated with mainstream iterative optimization methods for PINN training, i.e., Section~\ref{sec:results} equation (\ref{eq:SGD}) (see schematic in Fig.~\ref{fig:schemetic}).

The derivation of the sequential correction term begins by reformulating the PDE constraint in equation (\ref{eq:pde_eqn}) as an iterative update based on modified Richardson iteration (i.e., $\mathbb{B}= I$) under an intermediate iteration step $k$:
\begin{align}
    \label{eq:MRI}
    f(\cdot;\boldsymbol{w}^{k})-f(\cdot;\boldsymbol{w}^{k-1}) = \tau_{sc} \mathcal{R}
\end{align}
where $\mathcal{R}= h(\cdot)-\mathcal{N}_\vartheta[f(\cdot;\boldsymbol{w}^{k})]$ is the residual and $\tau_{sc}$ is the hyperparameter. We introduce the auxiliary function $\mathcal{F}(= f(\cdot;\boldsymbol{w}^{k})-f(\cdot;\boldsymbol{w}^{k-1}))$ that modifies the PDE loss at each optimization iteration to help prevent premature convergence in a poor local minima. Drawing inspiration from the implicit residual smoothing method, equation (\ref{eq:MRI}) can be casted as: 
\begin{align}
    \label{eq:IRS_fn}
    f(\cdot;\boldsymbol{w}^{k})-f(\cdot;\boldsymbol{w}^{k-1}) = \tau_{sc} \mathcal{\overline{R}}
\end{align}
$\mathcal{\overline{R}} = \Gamma * \mathcal{R}$ is a smoothed residual obtained by performing convolution operator $\Gamma$, such that it can provide sufficient smoothing to improve convergence while being computationally efficient. The following numerical convolution operator is chosen in this study:
\begin{align}
    \label{eq:gamma}
    \Gamma \approx \Gamma_\alpha = (I - {\alpha^2} \nabla^2 )^{-1}
\end{align}
where $\alpha$ is filtered length and acts as a hyperparameter.

It is noted that the above function is associated with Green function of Helmholtz equation, which has been utilized in Leray-$\alpha$ turbulence model~\cite{bib:Cheskidov04} as well as a smoothing kernel~\cite{bib:Chiu11,bib:Chiu23}. It has also been shown that based on choice of filtered length, $\alpha$, this function has the ability to filter the corresponding wavenumber~\cite{bib:BIENNER24,bib:Cheskidov04,bib:Ilyin06}, so as to improve convergence in typical PDE solvers. By substituting equation (\ref{eq:gamma}) into equation (\ref{eq:IRS_fn}), and defining residual smoothing operator $\mathcal{P}\equiv\Gamma^{-1}$, the following equations can be derived:
\begin{subequations}
    \begin{align}
        \frac{1}{\tau_{sc}}\mathcal{F} = \Gamma_\alpha * \mathcal{R} = (I - {\alpha^2} \nabla^2 )^{-1}\mathcal{R}\\
        \label{eq:IRS-final}
        \frac{1}{\tau_{sc}}\mathcal{P}_{\alpha}~\mathcal{F} = \mathcal{R}
    \end{align}
\end{subequations}
where $\mathcal{P}_{\alpha} = (I - {\alpha^2} \nabla^2 )$. When the solution is fully converged, the auxiliary term $\mathcal{F}\left(=f(\cdot;\boldsymbol{w}^k) - f(\cdot;\boldsymbol{w}^{k-1})\right)$ will vanish, ensuring the PDE residual $\mathcal{R}\left(= h(\cdot)-\mathcal{N}_\vartheta[f(\cdot;\boldsymbol{w}^{k})] \right)$ still equates to zero.

We employ the above equation (\ref{eq:IRS-final}) as the sequential corrected PDE loss function $\mathcal{L}_\textit{sc-pde}$ at iteration $k >0$,
\begin{align}
    \label{eq:SC_pde}
    \mathcal L_\textit{sc-pde}^k = \Vert &\mathcal{N}_\vartheta[f(\cdot;\boldsymbol{w}^k)] - h(\cdot)+ \frac{1}{\tau_{sc}}\mathcal{P}_{\alpha}~\left[ f(\cdot;\boldsymbol{w}^{k})-f(\cdot;\boldsymbol{w}^{k-1}) \right]  \Vert _{L^2(\Omega \times (0,T])}^2
\end{align}
Algorithm~\ref{algo:scale-pinn} summarizes the computational procedures for the present Scale-PINN. In summary, the algorithm: (1) infuses the concept of numerical algorithms with enhanced robustness and stability to improve convergence; (2) converges to the original system when the loss value approaches zero, so it does not affect the ultimate accuracy; and (3) remains simple and easy to implement so as to ensure that the overall computational time does not increase significantly.

\begin{algorithm}
    \caption{Sequential Correction Algorithm for Learning Efficient PINN (Scale-PINN) \label{algo:scale-pinn}}
    \begin{flushleft}
            \textbf{INPUT:} network architecture $f$, initial network weights $\boldsymbol{w}^0$, $\boldsymbol{w}^{-1} = \boldsymbol{w}^0$ and learning algorithm hyperparameters ($\tau_{sc}$, $\tau_{\alpha}$, $\gamma$, $\lambda_{ic}$, $\lambda_{bc}$, $\eta$)\\
            \textbf{OUTPUT:} $f(\cdot;\boldsymbol{w})$
    \end{flushleft}
    \begin{algorithmic}[1] 
        \For{$k=0,...,N$}
            \State Compute the loss terms $\mathcal L_{ic}$ and $\mathcal L_{bc}$ through equation~\ref{eq:pinn_ic_fn} and equation~\ref{eq:pinn_bc_fn}:
    \begin{align} 
            \nonumber
            \mathcal{L}_{ic}^k &= \lVert f(\cdot,t=0;\boldsymbol{w}^k) - u_0(\cdot) \rVert _{L^2(\Omega)}^2 \\ 
            \nonumber
            \mathcal{L}_{bc}^k &= \lVert \mathcal{B}[f(\cdot;\boldsymbol{w}^k)] - g(\cdot) \rVert _{L^2(\partial\Omega \times (0,T])}^2
    \end{align}
            \State Compute the sequential corrected PDE loss term $\mathcal L_\textit{sc-pde}^k$ by equation~\ref{eq:SC_pde}:
            \begin{align}
                \nonumber
                \mathcal L_\textit{sc-pde}^k &= \Vert \mathcal{N}_\vartheta[f(\cdot;\boldsymbol{w}^k)] - h(\cdot)+ \frac{1}{\tau_{sc}}\mathcal{P}_{\alpha}~\mathcal{F}  \Vert _{L^2(\Omega \times (0,T])}^2 \\ 
                \nonumber
                 \frac{1}{\tau_{sc}}\mathcal{P}_{\alpha}~\mathcal{F} &= \frac{1}{\tau_{sc}}\left((f(\cdot;\boldsymbol{w}^k) - f(\cdot;\boldsymbol{w}^{k-1})\right)-\frac{\gamma}{\tau_{\alpha}}\left(\nabla^2f(\cdot;\boldsymbol{w}^k)
                - \nabla^2f(\cdot;\boldsymbol{w}^{k-1})\right) 
            \end{align}
            \State Compute the Scale-PINN objective function:
            \begin{align}
                \nonumber
                \mathcal{L}_{sc}(\boldsymbol{w}^{k}) = \mathcal{L}_\textit{sc-pde}^k + \lambda_{ic}\ \mathcal{L}_{ic}^k + \lambda_{bc}\ \mathcal{L}_{bc}^k
                \end{align}
                \State Update the parameters $\boldsymbol{w}$ via gradient descent (such as SGD and Adam algorithms) with learning rate $\eta$:
                \begin{align}
                \nonumber
                \boldsymbol{w}^{k+1} = \boldsymbol{w}^{k} - \eta \nabla \mathcal{L}_{sc}(\boldsymbol{w}^{k})
            \end{align}
            %\State Check if $|\mathcal{L}_{SC}(\boldsymbol{w}^{k})| < \epsilon$
        \EndFor
    \end{algorithmic}
\end{algorithm}

Other choices of the residual smoothing operator $\mathcal{P}$ can be made based on prior knowledge, domain expertise, and desired convergence properties specific to the physics. For example, bi-Laplacian operator has been chosen in~\cite{bib:BIENNER24}, resulting in solutions with better accuracy. It is also noted that equation (\ref{eq:MRI}) is a special case of equation (\ref{eq:IRS_fn}), when the identity operator $\mathcal{I}$ is chosen as the residual smoothing operator ($\mathcal{P} = \mathcal{P}_I = I$).

\subsection{Description of the PINN simulation problem} \label{sec:PINN-simulation}

\subsubsection{Incompressible Navier-Stokes equations} \label{sec:ns-equation}
Under the isothermal and steady-state assumption, the incompressible N-S equations that govern the fluid flows can be expressed as continuity and momentum equations:
\begin{subequations}
    \label{eq:NS2}
    \begin{align}
        \label{eq:continuity2}
        \nabla\cdot \vec{u} &= 0 \\
        \label{eq:momentum2}
        (\vec{u}\cdot\nabla)\vec{u} &= \frac{1}{Re}\nabla^2 \vec{u} - \nabla p
    \end{align}
\end{subequations}
The dependent variables $\vec{u}=[u,v]^\intercal$ represent the velocity, and $p$ represents the pressure. The non-dimensional parameter $Re$ represents the ratio between inertial forces and viscous forces. Complex physical phenomenon can be observed with an increased Reynolds number ($Re$). The N-S equations are notoriously difficult to accurately solve, due to the high nonlinearity, convection instability and the strict constraint of mass conservation.

The sequential corrected PDE loss terms are derived in Section~\ref{sec:results} equations (\ref{eq:momentum-scale}-\ref{eq:continuity-scale}), with $\gamma$ set as $1/Re$. We fine-tune $\tau_{sc}$ and $\tau_\alpha$ for each problem below.

\textbf{Lid-driven cavity flow problems.} We simulate the lid-driven cavity flow with the top lid velocity ($u_{lid} = 1$) inside a 2D unit square, $x \in [0,1]$, $y \in [0,1]$, from $Re=400$ to $Re=20k$. To validate the Scale-PINN results, we generate high-fidelity reference solutions using the coupled version of improved divergence-free-condition compensated coupled (IDFC$^2$) method \cite{bib:Chiu21}, based on the quasi multi-moment framework and dispersion-relation preserving finite volume convection scheme. Ansys Fluent is also employed to generate the simulation results for time comparisons of the lid-driven cavity problem with $Re=3200$. To ensure the accuracy of the solution, QUICK scheme~\cite{bib:Leonard79} is chosen for the convection term, while the SIMPLE method~\cite{bib:Patankar80} is utilized for the velocity-pressure coupling. For generating reference solution of lid-driven cavity flow problem with $Re=20k$, a pseudo-transient coupled solver is employed to ensure convergence. The resultant linear system is solved by the algebraic multi-grid solver. These reference solutions, under mesh resolution of $512\times 512$ as ground truth (mesh-independence tests confirm that a $512 \times 512$ mesh provides a converged ground-truth solution), are then down-sampled to 100$\times$100 ($Re=400,3200$), 150$\times$150 ($Re=7500$), and 256$\times$256 ($Re=10k,20k$) sample points for the PINN validation.

\textbf{Flow past obstacles problems.} Three scenarios of flow past obstacles have been investigated in this study to validate the applicability and efficiency of Scale-PINN:
\begin{enumerate}[(1)]
    \item
    Single NACA0012 airfoil with $7^\circ$ angle of attack, $Re=1000$
    \item
    Two-staggered NACA0012 airfoils with $7^\circ$ angle of attack, $Re=500$
    \item
    Three-staggered square cylinders, $Re=25$
\end{enumerate}
The above canonical problems are relevant for engineering applications in aerodynamics and urban flow. The computational domain for the single airfoil scenario is $x \in [-3,5]$, $y \in [-2,2]$. For the two staggered airfoils case, the horizontal and vertical distances between the two staggered airfoils are 0.5 and 0.2 respectively, in the domain $x \in [-3,7]$, $y \in [-2.5,2.5]$. For the three staggered square cylinders case (a canonical proxy for wind flow around buildings), three unit squares are located at $(8,-2)$, $(10,2)$, $(12,0)$ in a domain defined by $x \in [0,30]$, $y \in [-7.5,7.5]$. For all three flow past obstacles problems, uniform inlet is employed at left boundary ($u=1,v=0$), pressure outlet is employed at right boundary ($p=0$), while the slip boundary condition is employed for side boundaries ($\frac{\partial u}{\partial y}=0, v=0$). 

IDFC$^2$, together with the convolutional direct forcing immersed boundary (cDFIB) method \citep{bib:Chiu23}, is employed to generate high-fidelity reference solutions with the complex geometries, under the mesh resolutions of $2048\times1024$, $2560\times 1280$, and $1024\times512$, respectively, for single-airfoil, two-staggered airfoils, and three-staggered square cylinders. These reference solutions are then down-sampled to 801$\times$401 (single-airfoil) and 1001$\times$501 (two-staggered airfoils and three-staggered square cylinders) sample points for the PINN validation.

\textbf{Rayleigh-B\'{e}nard convection problem.} Rayleigh-B\'{e}nard convection is a thermal instability phenomenon due to the temperature difference between the bottom hot plane and the top cold plane~\cite{bib:Castaing1989RB}. When the buoyancy forces overcome the viscous forces, flow starts to develop and result in convection cells, and can lead to transient and chaotic behavior when the temperature-gradient-driven buoyancy forces dominates. The problem is particularly relevant in urban sustainability studies, where modeling natural convection processes can inform energy efficiency, ventilation, and thermal comfort assessment.

The governing equations of the multiphysics transient dynamics can be written as follows:
\begin{subequations} \label{eq:RB}
    %\small
    \begin{align}
        &\frac{\partial u}{\partial x} + \frac{\partial v}{\partial y} = 0 \label{eq:RB-div} \\ 
        &\frac{\partial u}{\partial t} + u \frac{\partial u}{\partial x} + v \frac{\partial u}{\partial y} 
        = \sqrt{\frac{Pr}{Ra}} \left( \frac{\partial^2 u}{\partial x^2} + \frac{\partial^2 u}{\partial y^2} \right) - \frac{\partial p}{\partial x} \label{eq:RB-m1} \\ 
        &\frac{\partial v}{\partial t} + u \frac{\partial v}{\partial x} + v \frac{\partial v}{\partial y} 
        = \sqrt{\frac{Pr}{Ra}} \left( \frac{\partial^2 v}{\partial x^2} + \frac{\partial^2 v}{\partial y^2} \right) - \frac{\partial p}{\partial y} + T \label{eq:RB-m2} \\ 
        &\frac{\partial T}{\partial t} + u \frac{\partial T}{\partial x} + v \frac{\partial T}{\partial y} 
        = \frac{1}{\sqrt{Pr\,Ra}} \left( \frac{\partial^2 T}{\partial x^2} + \frac{\partial^2 T}{\partial y^2} \right) \label{eq:RB-T}
    \end{align}
\end{subequations}
In the above, the dependent variables $\vec{u}=[u,v]^\intercal$ represent the velocity, $p$ represents the pressure, and $T$ represents the temperature. $Ra$ is the Rayleigh number that describes the ratio between buoyancy forces and viscous forces, and $Pr$ is the Prandtl number that represents the ratio between momentum diffusivity and thermal diffusivity.

We simulate the transient dynamic in a spatial domain $x \in [0,4]$, $y \in [0,1]$, and time domain $t \in [0,50]$, with $Ra = 100k$ and $Pr = 0.71$, and boundary conditions $T = 0.5$ for the bottom hot plane and $T = -0.5$ for the top cold plane. The side boundary conditions are set as adiabatic. In this study, the steady-state solution with $Ra = 2k$ is used as initial condition to ensure the uniqueness of the transient behavior (the problem is sensitive to the initial condition~\cite{bib:Soong96RB,bib:Li2012RB}). The reference solution is generated under mesh resolution of $1024\times 256$ with time step size $0.001$ by second order backward differentiation formula using IDFC$^2$ solver. This reference solution is then down-sampled to $258\times 66\times 501$ spatio-temporal sample points for PINN validation.

The Scale-PINN objective function for simulating the Rayleigh-B\'{e}nard convection is thus defined as: $\mathcal{L}_\textit{sc}(\boldsymbol{w}^k) = \mathcal L_\textit{sc-pde(Rc)}^k + \mathcal L_\textit{sc-pde(Ru)}^k + \mathcal L_\textit{sc-pde(Rv)}^k + \mathcal L_\textit{sc-pde(RT)}^k + \lambda_{bc} \mathcal{L}_{bc}^k + \lambda_{ic} \mathcal{L}_{ic}^k$. We derive the sequential corrected PDE loss terms for equations (\ref{eq:RB-div}-\ref{eq:RB-T}) as:
\begin{subequations}
    \label{eq:RB-scale}
    \small
    \begin{align}
        \mathcal L_\textit{sc-pde(Rc)}^k &= \Vert \frac{\partial u^k}{\partial x} + \frac{\partial v^k}{\partial y} + \mathcal S_\textit{Rc} \Vert_{L^2(\Omega)}^2 \\  
        \mathcal L_\textit{sc-pde(Ru)}^k &= \Vert \frac{\partial u^k}{\partial t} + u^k\frac{\partial u^k}{\partial x} + v^k\frac{\partial u^k}{\partial y} -\sqrt{\frac{Pr}{Ra}}(\frac{\partial^2 u^k}{\partial x^2} + \frac{\partial^2 u^k}{\partial y^2}) + \frac{\partial p^k}{\partial x} + \mathcal S_\textit{Ru} \Vert_{L^2(\Omega)}^2 \\
        \mathcal L_\textit{sc-pde(Rv)}^k &= \Vert \frac{\partial v^k}{\partial t} + u^k\frac{\partial v^k}{\partial x} + v^k\frac{\partial v^k}{\partial y} -\sqrt{\frac{Pr}{Ra}}(\frac{\partial^2 v^k}{\partial x^2} + \frac{\partial^2 v^k}{\partial y^2}) + \frac{\partial p^k}{\partial y} + \mathcal S_\textit{Rv} \Vert_{L^2(\Omega )}^2 \\
        \mathcal L_\textit{sc-pde(RT)}^k &= \Vert \frac{\partial T^k}{\partial t} + u^k\frac{\partial T^k}{\partial x} + v^k\frac{\partial T^k}{\partial y} -\sqrt{\frac{1}{PrRa}}(\frac{\partial^2 T^k}{\partial x^2} + \frac{\partial^2 T^k}{\partial y^2}) + \mathcal S_\textit{RT} \Vert_{L^2(\Omega )}^2 \\   
        \mathcal S_\textit{Rc} &= \frac{1}{\tau_{sc}}(p^k-p^{k-1}) \\
        \mathcal S_\textit{Ru} &= \frac{1}{\tau_{sc}}(u^k-u^{k-1}) -\frac{\gamma_{Ruv}}{\tau_{\alpha}}\left[ (\frac{\partial^2 u^k}{\partial x^2} + \frac{\partial^2 u^k}{\partial y^2})-(\frac{\partial^2 u^{k-1}}{\partial x^2} + \frac{\partial^2 u^{k-1}}{\partial y^2}) \right] \\
        \mathcal S_\textit{Rv} &= \frac{1}{\tau_{sc}}(v^k-v^{k-1}) -\frac{\gamma_{Ruv}}{\tau_{\alpha}}\left[ (\frac{\partial^2 v^k}{\partial x^2} + \frac{\partial^2 v^k}{\partial y^2})-(\frac{\partial^2 v^{k-1}}{\partial x^2} + \frac{\partial^2 v^{k-1}}{\partial y^2}) \right] \\
        \mathcal S_\textit{RT} &= \frac{1}{\tau_{sc}}(T^k-T^{k-1}) -\frac{\gamma_{RT}}{\tau_{\alpha}}\left[ (\frac{\partial^2 T^k}{\partial x^2} + \frac{\partial^2 T^k}{\partial y^2})-(\frac{\partial^2 T^{k-1}}{\partial x^2} + \frac{\partial^2 T^{k-1}}{\partial y^2}) \right]
    \end{align}
\end{subequations}
where $\gamma_{Ruv} = \sqrt\frac{Pr}{Ra}$, $\gamma_{RT} = \sqrt{\frac{1}{PrRa}}$ (same as the diffusion coefficient), while $\tau_{sc}$ and $\tau_\alpha$ are separate tuning hyperparameters.

\subsubsection{Kuramoto-Sivashinsky equation} \label{sec:ks-equation}

The Kuramoto-Sivashinsky equation is a fourth order nonlinear PDE that models fluid film flows~\cite{bib:Kalogirou15}:
\begin{align}
    \frac{\partial u}{\partial t} + a_1 ~u\frac{\partial u}{\partial x} + a_2 \frac{\partial^2 u}{\partial x^2} + a_3\frac{\partial^4 u}{\partial x^4} = 0
\end{align}
Due to the interaction between the nonlinear term with the diffusion and anti-diffusion terms, the solutions of Kuramoto-Sivashinsky equation exhibit chaotic spatio-temporal patterns. We apply Scale-PINN to solve for the solution in spatial domain $x \in [0,2\pi]$ and time domain $t \in [0,0.4]$ with periodic spatial boundary condition and initial condition $u_0(x) = \cos(x)(1+\sin(x))$, where $a_1=\frac{100}{16}$, $a_2=\frac{100}{16^2}$ and $a_3=\frac{100}{16^2}$. The above settings make the system very stiff, and intrinsically hard to solve by a PINN model. The reference (512$\times$101) solution is obtained from~\cite{wang2023experts}, generated using the Chebfun package~\cite{bib:Driscoll14} employing a fourth-order stiff time-stepping scheme (ETDRK4)~\cite{bib:Cox02}.

The Scale-PINN objective function for simulating Kuramoto-Sivashinsky solution is thus defined as: $\mathcal{L}_\textit{sc}(\boldsymbol{w}^k) = \mathcal L_\textit{sc-pde(KS)}^k + \lambda_{bc} \mathcal{L}_{bc}^k + \lambda_{ic} \mathcal{L}_{ic}^k$. We derive the sequential corrected PDE loss for Kuramoto-Sivashinsky equation:
\begin{subequations}
    \label{eq:KS-scale}
    \small
    \begin{align}
            \mathcal L_\textit{sc-pde(KS)}^k &= \Vert \frac{\partial u^k}{\partial t} + a_1 ~u^k\frac{\partial u^k}{\partial x} + a_2 \frac{\partial^2 u^k}{\partial x^2} + a_3\frac{\partial^4 u^k}{\partial x^4} + \mathcal S_\textit{KS} \Vert_{L^2(\Omega)}^2 \\        
            \mathcal S_\textit{KS} &= \frac{1}{\tau_{sc}}(u^k-u^{k-1}) -\frac{\gamma_{KS}}{\tau_{\alpha}}\left[ (\frac{\partial^2 u^k}{\partial x^2} + \frac{\partial^2 u^k}{\partial y^2})-(\frac{\partial^2 u^{k-1}}{\partial x^2} + \frac{\partial^2 u^{k-1}}{\partial y^2}) \right] 
    \end{align}
\end{subequations}
We set $\gamma_{KS}=a_2$, which is the same magnitude of the anti-diffusion coefficient, while $\tau_{sc}$ and $\tau_\alpha$ are then fine-tuned.

\subsubsection{Grey-Scott equations} \label{sec:gs-equations}

The Gray–Scott equations describe nonlinear chemical kinetics governed by coupled reaction–diffusion dynamics~\cite{bib:Gray94}:
\begin{subequations}
    \begin{align}
        \frac{\partial u^k}{\partial t} &= \epsilon_1 (\frac{\partial^2 u}{\partial x^2} + \frac{\partial^2 u}{\partial y^2}) + b_1(1-u) - c_1 uv^2 \\
        \frac{\partial v}{\partial t} &= \epsilon_2 (\frac{\partial^2 v}{\partial x^2} + \frac{\partial^2 v}{\partial y^2}) - b_2v + c_2 uv^2
    \end{align}
\end{subequations}
We follow the settings in~\cite{wang2024piratenets}, i.e., $\epsilon_1 = 0.2$, $\epsilon_2 = 0.1$, $b_1 = 40$, $b_2 = 100$ and $c_1=c_2 = 1000$, with periodic spatial boundary condition. We apply Scale-PINN to solve for the solution in the spatio-temporal domain, $x \in [-1,1]$, $y \in [-1,1]$, and $t \in [0,0.5]$, given the initial condition:
\begin{subequations}
    \begin{align}
        u_0(x,y)& = 1-\exp(-10\left( (x+0.05)^2 + (y+0.02)^2 \right)\\
        v_0(x,y)& =\exp(-10\left( (x-0.05)^2 + (y-0.02)^2 \right)
    \end{align}
\end{subequations}
%With periodic spatial boundary condition, in this study the training is performed with 5 individual time windows.
The reference (200$\times$200$\times$26) solution is obtained from~\cite{wang2024piratenets}, generated with the ETDRK4 numerical scheme using Chebfun package.

The Scale-PINN objective function for simulating Gray–Scott solutions is thus defined as: $\mathcal{L}_\textit{sc}(\boldsymbol{w}^k) = \mathcal L_\textit{sc-pde(GSu)}^k + \mathcal L_\textit{sc-pde(GSv)}^k + \lambda_{bc} \mathcal{L}_{bc}^k + \lambda_{ic} \mathcal{L}_{ic}^k$. We derive the sequential corrected PDE loss for Gray–Scott equations:
\begin{subequations}
    \label{eq:GS-scale}
    \small
    \begin{align}
            \mathcal L_\textit{sc-pde(GSu)}^k &= \Vert \frac{\partial u^k}{\partial t} - \epsilon_1 (\frac{\partial^2 u^k}{\partial x^2} + \frac{\partial^2 u^k}{\partial y^2}) - b_1(1-u^k) + c_1 u^k(v^k)^2 + \mathcal S_\textit{GSu} \Vert_{L^2(\Omega)}^2 \\
            \mathcal L_\textit{sc-pde(GSv)}^k &= \Vert \frac{\partial v^k}{\partial t} - \epsilon_2 (\frac{\partial^2 v^k}{\partial x^2} + \frac{\partial^2 v^k}{\partial y^2}) + b_2v^k - c_2 u^k(v^k)^2 + \mathcal S_\textit{GSv} \Vert_{L^2(\Omega)}^2 \\
            \mathcal S_\textit{GSu} &= \frac{1}{\tau_{sc}}(u^k-u^{k-1}) -\frac{\gamma_{GSu}}{\tau_{\alpha}}\left[ (\frac{\partial^2 u^k}{\partial x^2} + \frac{\partial^2 u^k}{\partial y^2})-(\frac{\partial^2 u^{k-1}}{\partial x^2} + \frac{\partial^2 u^{k-1}}{\partial y^2}) \right] \\
            \mathcal S_\textit{GSv} &= \frac{1}{\tau_{sc}}(v^k-v^{k-1}) -\frac{\gamma_{GSv}}{\tau_{\alpha}}\left[ (\frac{\partial^2 v^k}{\partial x^2} + \frac{\partial^2 v^k}{\partial y^2})-(\frac{\partial^2 v^{k-1}}{\partial x^2} + \frac{\partial^2 v^{k-1}}{\partial y^2}) \right]
    \end{align}
\end{subequations}
We set $\gamma_{GSu}=\epsilon_1$, $\gamma_{GSv}=\epsilon_2$, while $\tau_{sc}$ and $\tau_\alpha$ are then fine-tuned.

\subsubsection{Korteweg–De Vries equation} \label{sec:kdv-equation}

The Korteweg–De Vries equation is a third order nonlinear dispersive PDE that models shallow water waves~\cite{bib:Korteweg1895}:
\begin{align}
    \frac{\partial u}{\partial t} + u\frac{\partial u}{\partial x} + \nu \frac{\partial^3 u}{\partial x^3} = 0
\end{align}
We apply Scale-PINN to solve for the solution for $\nu=(\frac{11}{500})^2$ in spatial domain $x \in [-1,1]$ and time domain $t \in [0,1]$ with periodic spatial boundary condition and initial condition $u_0(x) = \cos(\pi x)$. The reference (512$\times$201) solution is obtained from~\cite{wang2024piratenets}, generated with the ETDRK4 numerical scheme using Chebfun package.

The Scale-PINN objective function for simulating Korteweg–De Vries solution is thus defined as: $\mathcal{L}_\textit{sc}(\boldsymbol{w}^k) = \mathcal L_\textit{sc-pde(KdV)}^k + \lambda_{bc} \mathcal{L}_{bc}^k + \lambda_{ic} \mathcal{L}_{ic}^k$. We derive the sequential corrected PDE loss for Korteweg–De Vries equation:
\begin{subequations}
    \label{eq:KdV-scale}
    \small
    \begin{align}
            \mathcal L_\textit{sc-pde(KdV)}^k &= \Vert \frac{\partial u^k}{\partial t} + u^k\frac{\partial u^k}{\partial x} + \nu \frac{\partial^3 u^k}{\partial x^3} + \mathcal S_\textit{KdV} \Vert_{L^2(\Omega)}^2 \\        
            \mathcal S_\textit{KdV} &= \frac{1}{\tau_{sc}}(u^k-u^{k-1}) -\frac{\gamma_{KdV}}{\tau_{\alpha}}\left[ (\frac{\partial^2 u^k}{\partial x^2} + \frac{\partial^2 u^k}{\partial y^2})-(\frac{\partial^2 u^{k-1}}{\partial x^2} + \frac{\partial^2 u^{k-1}}{\partial y^2}) \right] 
    \end{align}
\end{subequations}
We set $\gamma_{KdV}=\sqrt{\nu}$, which is the square-root of the dispersion coefficient, while $\tau_{sc}$ and $\tau_\alpha$ are then fine-tuned.

\subsubsection{Allen-Cahn equation} \label{sec:ac-equation}

Both Korteweg–De Vries and Allen–Cahn equations are commonly studied benchmark problems in the PINN literature. The Allen–Cahn equation models crystal growth and phase separation as a diffusion–reaction process~\cite{bib:ALLEN1975}:
\begin{align}
    \frac{\partial u}{\partial t} - \alpha\frac{\partial^2 u}{\partial x^2} + \delta~(u^3-u) = 0
\end{align}
We apply Scale-PINN to solve for the solution for $\alpha = 0.0001$ and $\delta = 5$ in spatial domain $x \in [-1,1]$ and time domain $t \in [0,1]$ with periodic spatial boundary condition and initial condition $u_0(x) = x^2 \cos(\pi x)$. The reference (512$\times$201) solution is obtained from~\cite{wang2024piratenets}, generated with the ETDRK4 numerical scheme using Chebfun package.

The Scale-PINN objective function for simulating Allen–Cahn solution is thus defined as: $\mathcal{L}_\textit{sc}(\boldsymbol{w}^k) = \mathcal L_\textit{sc-pde(AC)}^k + \lambda_{bc} \mathcal{L}_{bc}^k + \lambda_{ic} \mathcal{L}_{ic}^k$. We derive the sequential corrected PDE loss for Allen–Cahn equation:
\begin{subequations}
    \label{eq:AC-scale}
    \small
    \begin{align}
            \mathcal L_\textit{sc-pde(AC)}^k &= \Vert \frac{\partial u^k}{\partial t} - \alpha\frac{\partial^2 u^k}{\partial x^2} + \delta~((u^k)^3-u^k) + \mathcal S_\textit{AC} \Vert_{L^2(\Omega)}^2 \\        
            \mathcal S_\textit{AC} &= \frac{1}{\tau_{sc}}(u^k-u^{k-1}) -\frac{\gamma_{AC}}{\tau_{\alpha}}\left[ (\frac{\partial^2 u^k}{\partial x^2} + \frac{\partial^2 u^k}{\partial y^2})-(\frac{\partial^2 u^{k-1}}{\partial x^2} + \frac{\partial^2 u^{k-1}}{\partial y^2}) \right] 
    \end{align}
\end{subequations}
In the above, $\gamma_{AC}$ is set as $\alpha$.
$\tau_{sc}$ and $\tau_\alpha$ are then fine-tuned.

\subsection{Scale-PINN model architecture and training strategies} \label{sec:PINN-full}

\subsubsection{Neural architecture and activation function design}

Scale-PINN employs a multi-layer perceptron (MLP) architecture as the backbone network, chosen for its proven effectiveness in approximating dynamical process, as well as its flexible design and ease of implementation. To more effectively learn a model output---mapped from the spatio-temporal input coordinates---that captures high-frequency features, which are prevalent in many dynamical systems, we initialize the network with artificial high-frequency components by modulating the first hidden layer with a factor of $F \pi$ in combination with a sine activation, as illustrated in Fig. \ref{fig:schemetic}. Here, $F$ serves as a problem-specific tuning parameter that controls the initial high-frequency range. During training, these frequencies are naturally reduced to an appropriate range\textemdash a process we refer to as \textit{frequency annealing}.

Two specialized MLP architectures are designed to accommodate the characteristics of different PDE problems.

\textbf{N-S flow network.} The network consists of multiple shared hidden layers mapped from spatio-temporal input coordinates, which then branch into variable-specific hidden layers for $u$, $v$, and $p$, respectively. For Rayleigh-B\'{e}nard convection problem, the network contains an additional branch for $T$. 
\begin{subequations}
    \small
    \nonumber
    \begin{align}
        f_{u}(x, t;\boldsymbol{w}) &= \boldsymbol{W}_{u,L} \mathbf{x}_{u,L} & \text{(\textit{output layer: u})} \\
        \mathbf{x}_{u,L} &= \psi(\boldsymbol{W}_{u,L-1} \mathbf{x}_{u,L-1} + \mathbf{b}_{u,L-1}) \\
        &\vdots \notag \\
        f_{v}(x, t;\boldsymbol{w}) &= \boldsymbol{W}_{v,L} \mathbf{x}_{v,L} & \text{(\textit{output layer: v})} \\
        \mathbf{x}_{v,L} &= \psi(\boldsymbol{W}_{v,L-1} \mathbf{x}_{v,L-1} + \mathbf{b}_{v,L-1}) \\
        &\vdots \notag \\
        f_{p}(x, t;\boldsymbol{w}) &= \boldsymbol{W}_{p,L} \mathbf{x}_{p,L} & \text{(\textit{output layer: p})} \\
        \mathbf{x}_{p,L} &= \psi(\boldsymbol{W}_{p,L-1} \mathbf{x}_{p,L-1} + \mathbf{b}_{p,L-1}) \\
        &\vdots \notag \\
        \mathbf{x}_{u,1}, \mathbf{x}_{v,1}, \mathbf{x}_{p,1} &\gets \mathbf{x}_{L}  & \text{(\textbf{multi-branch})} \\
        \mathbf{x}_{L} &= \psi(\boldsymbol{W}_{L-1} \mathbf{x}_{L-1} + \mathbf{b}_{L-1}) & \text{(\textit{shared hidden layers})} \\ 
        &\vdots \notag \\
        \mathbf{x}_{3} &= \psi(\mathbf{z}_{2}) & \text{(\textit{after activation})} \\     
        \mathbf{z}_{2} &= \boldsymbol{W}_2 \mathbf{x}_2 + \mathbf{b}_2 & \text{(\textit{2nd hidden layer})} \\ 
        \mathbf{x}_{2} &= \text{sin}(\mathbf{z}_{1}) & \text{(\textbf{frequency annealing})} \\     
        \mathbf{z}_{1} &\gets F \pi \mathbf{z}_{1} & \text{(\textbf{frequency annealing})} \\
        \mathbf{z}_{1} &= \boldsymbol{W}_1 \mathbf{x}_1 + \mathbf{b}_1 & \text{(\textit{1st hidden layer})} \\
        \mathbf{x}_{1} &\equiv (x,t) & \text{(\textit{input})}
    \end{align}
\end{subequations} 
where $\boldsymbol{w} = [\boldsymbol{W}_1, \mathbf{b}_1, ..., \boldsymbol{W}_{L-1}, \mathbf{b}_{L-1}, ..., \boldsymbol{W}_{u,L-1}, \mathbf{b}_{u, L-1}, \boldsymbol{W}_{u,L}, ..., \boldsymbol{W}_{v,L}, ...,  \boldsymbol{W}_{p,L}]$. The N–S flow network uses the SiLU activation function starting from the second hidden layer. We note that N–S flow network is a robust and highly performant neural architecture for many N-S flow problems~\cite{bib:Chiu22, bib:Wong23, wei2025ffv}.

\textbf{Skip connections network.} The network utilizes concatenative skip connections such that all the nonlinear hidden layers are concatenated at the final hidden layer.
\begin{subequations}
    \small
    \nonumber
    \begin{align}
        f(x, t;\boldsymbol{w}) &= \boldsymbol{W}_L \mathbf{x}_L & \text{(\textit{output layer})} \\
        \mathbf{x}_L &\gets \text{concatenate}(\mathbf{x}_L, \mathbf{x}_{L-1}, ..., \mathbf{x}_2)  & \text{(\textbf{skip connections})} \\
        \mathbf{x}_L &= \psi(\mathbf{z}_{L-1}) \\ 
        &\vdots \notag \\
        \mathbf{x}_{3} &= \psi(\mathbf{z}_{2}) & \text{(\textit{after activation})} \\     
        \mathbf{z}_{2} &= \boldsymbol{W}_2 \mathbf{x}_2 + \mathbf{b}_2 & \text{(\textit{2nd hidden layer})} \\ 
        \mathbf{x}_{2} &= \text{sin}(\mathbf{z}_{1}) & \text{(\textbf{frequency annealing})} \\     
        \mathbf{z}_{1} &\gets F \pi \mathbf{z}_{1} & \text{(\textbf{frequency annealing})} \\
        \mathbf{z}_{1} &= \boldsymbol{W}_1 \mathbf{x}_1 + \mathbf{b}_1 & \text{(\textit{1st hidden layer})} \\
        \mathbf{x}_1 &\equiv (x,t) & \text{(\textit{input})}
    \end{align}
\end{subequations} 
where $\boldsymbol{w} = [\boldsymbol{W}_1, \mathbf{b}_1, ..., \mathbf{W}_L]$. The skip connections network uses either the SiLU or softplus activation function starting from the second hidden layer. This neural architecture design allows us to increase the output layer width by stacking multiple hidden layers, thereby effectively improving the expressivity of the network, while maintaining a moderate number of nodes in each hidden layer. In addition, we note that the final hidden layer concatenation bears similarity to the way one constructs a polynomial basis space such as the monomial basis space. The additional operations at each hidden layer are analogous to the recurrence relations used in generating Chebyshev polynomials, and incorporation and concatenation of more hidden layers in the MLP essentially results in the creation of a larger (albeit finite) and more expressive basis space (with less truncation).

\subsubsection{Model training and hyperparameters}

Scale-PINN is trained using the Adam optimizer with a warm-up cosine decay learning rate schedule, where the minimum learning rate is set to $1e^{\text{-10}}$. Table~\ref{tab:summary} provides a summary of the Scale-PINN model architecture and training settings for all the studied PDE problems.

\subsubsection{Computational Environment}

All benchmark experiments are conducted on a workstation using a single NVIDIA GeForce RTX 3090 GPU. The Scale-PINN algorithm is implemented in the JAX framework to leverage its efficiency in automatic differentiation and linear algebra operations~\cite{bradbury2018jax,evojax2022}.

\backmatter

%\bmhead{Supplementary information}

%If your article has accompanying supplementary file/s please state so here. 

%Authors reporting data from electrophoretic gels and blots should supply the full unprocessed scans for key as part of their Supplementary information. This may be requested by the editorial team/s if it is missing.

%Please refer to Journal-level guidance for any specific requirements.

%\bmhead{Acknowledgements}

%Acknowledgements are not compulsory. Where included they should be brief. Grant or contribution numbers may be acknowledged.

%Please refer to Journal-level guidance for any specific requirements.

\bmhead{Code availability}

The example codes with instructions are available at \url{https://github.com/chiuph/SCALE-PINN}.

\clearpage

\begin{appendices}

\section{Extended Data}\label{secA1}

Table~\ref{tab:summary} provides a summary of the Scale-PINN model architecture and training settings for all the studied PDE problems.

%%%%%%%%%%%%%%%%%%%% TABLE A1 %%%%%%%%%%%%%%%%%%%%

\begin{sidewaystable}
\caption{Scale-PINN model architecture and training settings.} \label{tab:summary}
\begin{tabular*}{\textheight}{@{\extracolsep\fill}l>{\raggedright\arraybackslash}p{3.7cm}>{\raggedright\arraybackslash}p{5cm}>{\raggedright\arraybackslash}p{1.5cm}p{1.5cm}>{\raggedright\arraybackslash}p{1.2cm}>{\raggedright\arraybackslash}p{1.2cm}>{\raggedright\arraybackslash}p{1.2cm}>{\raggedright\arraybackslash}p{2cm}>{\raggedright\arraybackslash}p{2cm}}
\toprule%
& Problem & Neural architecture\footnotemark[1] & Frequency aneling, $[F]\pi$ & Activation & Batch size /iter. & No. training iter. & Initial learning rate & Loss function, $\lambda_{ic}$, $\lambda_{bc}$ & Sequential corrected loss term, $\tau_{sc}$, $\tau_\alpha$ \\
\midrule
1a & Navier-Stokes equations: lid-driven cavity flow $Re=400-3200$ & $(x,y)-\underline{128}-32-32-$ $[32-32-32-(u)$, $32-32-32-(v)$, $32-32-32-(p)]$ & $2\pi$ & silu & 400 & 50$k$ & $1e^{\text{-3}}$ - $5e^{\text{-4}}$ & {\footnotesize -, 10 - 15} & {\footnotesize 0.06 - 0.095, 0.5 - 1} \\[0.8cm]
1b & Navier-Stokes equations: lid-driven cavity flow $Re=7500-20k$ & $(x,y)-\underline{256}-64-64-$ $[64-64-64-(u)$, $64-64-64-(v)$, $64-64-64-(p)]$ & $2\pi$ & silu & 1,000 - 2,400 & 50$k$ - 100$k$ & $5e^{\text{-4}}$ & {\footnotesize -, 10 - 20} & {\footnotesize 0.095 - 0.11, 0.5 - 0.6} \\[0.8cm]
2 & Navier-Stokes equations: 1-NACA0012 airfoil & $(x,y)-\underline{64}-32-32-$ $[32-32-32-(u)$, $32-32-32-(v)$, $32-32-32-(p)]$ & $\pi$ & silu & 4,000 & 50$k$ & $5e^{\text{-3}}$ & {\footnotesize -, 5}  & {\footnotesize 0.1, 0.5} \\[0.5cm]
3 & Navier-Stokes equations: 2-staggered airfoils & $(x,y)-\underline{64}-32-32-$ $[32-32-32-(u)$, $32-32-32-(v)$, $32-32-32-(p)]$ & $\pi$ & silu & 4,000 & 50$k$ & $5e^{\text{-3}}$ & {\footnotesize -, 5}  & {\footnotesize 0.05, 1} \\[0.5cm]
4 & Navier-Stokes equations: 3-staggered square cylinders & $(x,y)-\underline{64}-32-32-$ $[32-32-32-(u)$, $32-32-32-(v)$, $32-32-32-(p)]$ & $\pi$ & silu & 4,000 & 80$k$ & $5e^{\text{-3}}$ & {\footnotesize -, 5} & {\footnotesize 0.035, 15} \\[0.5cm]
5 & Navier-Stokes equations: Rayleigh-B\'{e}nard convection & $(x,y,t)-\underline{128}-64-64-$ $[64-64-64-(u)$, $64-64-64-(v)$, $64-64-64-(p)$, $64-64-64-(T)]$ & $4\pi$ & silu & 4,000 & 50$k$ & $5e^{\text{-3}}$ & {\footnotesize 1, 10} & {\footnotesize 0.1, 1.5} \\[1cm]
6 & Kuramoto-Sivashinsky equation & $(x,t)-\underline{128}-128-128-128^{\ +}-(u)$ & $4\pi$ & silu & 1,000 & 200$k$ & $1e^{\text{-3}}$ & {\footnotesize 500, 5000} & {\footnotesize 0.2, 1.5} \\[0.5cm]
7 & Grey-Scott equations & $(x,y,t)-\underline{128}-128-128-128^{\ +}-(u,v)$ & $2\pi$ & silu & 1,000 & 300$k$ & $2e^{\text{-3}}$ & {\footnotesize 5000, 1000} & {\footnotesize 0.02, 10} \\[0.5cm]
8 & Korteweg–De Vries equation & $(x,t)-\underline{128}-128-128-128^{\ +}-(u)$ & $2\pi$ & softplus & 1,000 & 300$k$ & $1e^{\text{-3}}$ & {\footnotesize 1000, 1000} & {\footnotesize 0.1, 20} \\[0.5cm]
9 & Allen-Cahn equation & $(x,t)-\underline{128}-128-128-128-(u)$ & $2\pi$ & silu & 1,000 & 500$k$ & $2e^{\text{-3}}$ & {\footnotesize 100, 100} & {\footnotesize 0.4, 1.5} \\[0.1cm]
\botrule
\end{tabular*}
\footnotesize
\footnotetext[1]{For the MLP architecture, the numbers in between input and output represent the number of nodes in each hidden layer. For example, $(x)-\underline{64}-32-32-32^{\ +}-(u)$ indicates a single input $x$, followed by 4 hidden layers with 64, 32, 32 and 32 nodes in each layer, and a single output $u$. We apply the sinusoidal features mapping~\cite{wong2022learning} to replace \underline{first hidden layer} (frequency annealing) and initialize all network weights using \textit{He} method. Besides, the superscript $^{\ +}$ at final hidden layer indicates a concatenative skip connections such that all the nonlinear hidden layers are concatenated at the final hidden layer.}
\end{sidewaystable}

%%%%%%%%%%%%%%%%%%%% END OF TABLE A1 %%%%%%%%%%%%%%%%%%%%

\clearpage

\end{appendices}


\bibliography{sn-bibliography}% common bib file

%TC:endignore
% Start words count from now on!

\end{document}